\documentclass[%
 reprint,
superscriptaddress,
 amsmath,amssymb,
 aps,
prx,
]{revtex4-2}

\usepackage{mathtools}
\usepackage{graphicx}
\usepackage{xcolor}
\usepackage{enumerate}
\usepackage{wrapfig}
\usepackage{lipsum}  
\usepackage{dcolumn}
\usepackage{braket}
\usepackage{appendix}
\usepackage[colorlinks=True,citecolor={blue},linkcolor={blue},urlcolor={blue}]{hyperref} 
\usepackage{amsmath,amsthm,amssymb,amsfonts}
\usepackage{bm}
\usepackage{physics}

\begin{document}
\preprint{APS/123-QED}

\title{Purcell-Enhanced, Directional Light–Matter Interaction in a Waveguide-Coupled Nanocavity}

    \author{Nicholas~J.~Martin}
\email{n.j.martin@sheffield.ac.uk}
\affiliation{These authors contributed equally}
\affiliation{School of Mathematical and Physical Sciences, University of Sheffield, Sheffield S3 7RH, UK}%
\author{Dominic~Hallett}
\affiliation{These authors contributed equally}
\affiliation{School of Mathematical and Physical Sciences, University of Sheffield, Sheffield S3 7RH, UK}%
\author{Mateusz~Duda}
\author{Luke~Hallacy}
\author{Elena~Callus}
\author{Luke~Brunswick}
\author{René~Dost}
\affiliation{School of Mathematical and Physical Sciences, University of Sheffield, Sheffield S3 7RH, UK}%
\author{Edmund~Clarke}
\author{Pallavi~K.~Patil}
\affiliation{EPSRC National Epitaxy Facility, University of Sheffield, Sheffield S1 4DE, UK}%
\author{Pieter~Kok}
\author{Maurice~S.~Skolnick}
\author{Luke~R.~Wilson}

\affiliation{School of Mathematical and Physical Sciences, University of Sheffield, Sheffield S3 7RH, UK}%



\begin{abstract} 
We demonstrate electrically tunable, spin-dependent, directional coupling of single photons by embedding quantum dots (QDs) in a waveguide-coupled nanocavity. The directional behavior arises from direction-dependent interference between two cavity modes when coupled to the device waveguides. The small mode volume cavity enables simultaneous Purcell enhancement (${10.8\pm0.7}$) and peak directional contrast (${88\pm1\%}$), exceeding current state-of-the-art waveguide-only systems. We also present a scattering matrix model for the transmission through this structure, alongside a quantum trajectory-based model for predicting the system's directionality, which we use to explain the observed asymmetry in directional contrast seen in QD devices. Furthermore, the nanocavity enables wide-range electrical tuning of the emitter's directional contrast. We present results showing precise tuning of a QD emission line from a directional contrast of ${2\%}$ to ${96\%}$. In combination, these characteristics make this cavity-waveguide approach promising for use as a building block in directional nanophotonic circuits.

\end{abstract}

\maketitle



\section{Introduction}
\begin{figure*}[t!]
\centering
\includegraphics[width=0.99\textwidth]{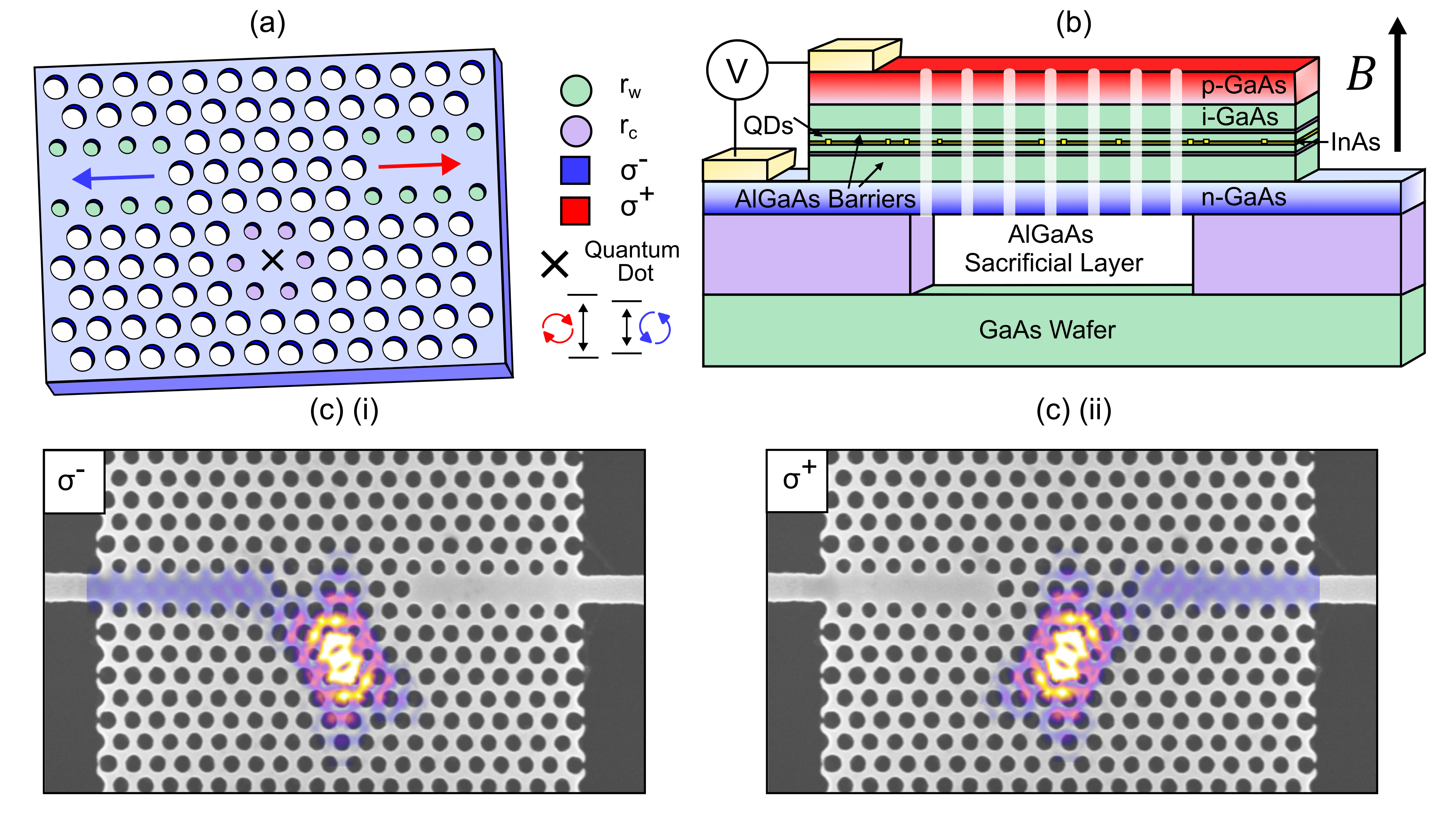}
\caption{(a) Schematic of the cavity design. (b) Stack diagram showing the structure of the QD wafer. (c) SEM image of the photonic crystal device overlaid with the time-averaged electric field intensity for the device with (i) a $\sigma^{-}$ and (ii) a $\sigma^{+}$ dipole at the center of the cavity. }
\label{fig:Design}
\end{figure*}

Achieving a high directional contrast alongside strong Purcell enhancement at the single-photon level has been a long-standing challenge in integrated quantum nanophotonics. Our novel nanocavity-waveguide system overcomes this barrier, advancing the potential for scalable quantum photonic technologies. Integrated nanophotonic systems have demonstrated the generation~\cite{Antoniadis2022,H1Cav2_2018}, manipulation~\cite{add_drop,Lodahl2004,Anisotropic_4,Volz2014}, and detection~\cite{Najafi2015,Gyger2021} of quantum states in a scalable, on-chip architecture. For example, a highly effective two-way conduit for light and matter can be realized through the integration of an intrinsic quantum emitter (QE) within a nanophotonic waveguide supporting a single optical mode. Notable examples of such QEs include semiconductor quantum dots (QDs)~\cite{PhysRevLett.113.093603}, diamond color centers~\cite{SiV4}, and single atoms~\cite{PhysRevLett.130.103601,Volz2014}.

Enhanced light–matter interaction, quantified by the Purcell factor ($F_P$), offers transformative benefits for quantum photonic devices. These include increased photon indistinguishability, higher coupling efficiencies, and faster device operation rates, making it foundational for scalable quantum technologies~\cite{PhysRev.69.37,Purcell2,PhysRevLett.113.093603,H1Cav2_2018, MD_switch}. By increasing the decay rate of a QE, the Purcell effect also reduces the impact of decoherence processes on the indistinguishability of emitted photons~\cite{Purcell1,Purcell2}.

Directional nanophotonic devices harness the coupling between QE spin states and photon propagation directions. By engineering the device geometry to create regions of circularly polarized electric fields, these systems can selectively interface with spin-dependent transitions of QEs. This control facilitates functionalities such as spin-to-path conversion~\cite{NB1_2016}, path-dependent spin initialization~\cite{NB2_2017}, and proposals for entangling multiple QEs~\cite{Gonzalez-Ballestro2015}. Additionally, the inherent nonreciprocity offers routes to compact optical circulators~\cite{doi:10.1126/science.aaj2118}.

Waveguide-based devices have made strides in achieving moderate Purcell factors (${~3<F_P<5}$) with robust directionality by embedding QEs in slow light regions of photonic crystal waveguides. However, these designs struggle to combine high Purcell enhancement and directionality due to spatial polarization variations inherent in the waveguide field~\cite{GP2_2017,hamidreza, PhysRevResearch.6.L022065}. Although highly directional emission from a QD has been achieved in waveguide structures~\cite{hamidreza}, the simultaneous combination of high directionality and large Purcell enhancement remains elusive, with the state of the art being
a Purcell factor of $F_P = 5 \pm 1$ for a chirally coupled QD~\cite{hamidreza}.

This work demonstrates a novel cavity-waveguide system that achieves a record combination of high Purcell enhancement (${F_P=10.8 \pm 0.7}$) and high directional contrast (${C=88 \pm 1\%}$), alongside an electrically adjustable directionality from $2\%$ to $96\%$. Unlike other approaches, our cavity-waveguide system can maintain a near-unity directional contrast even for emitter displacements up to $60$~nm from the cavity center, offering unprecedented robustness to fabrication imperfections~\cite{ChiralCav_2022}.

We develop theoretical models, including a scattering matrix model for transmission and a quantum trajectory-based model for directional contrast, to predict and optimize the system's performance. These models provide a deeper understanding of directionality and the interplay between system parameters, guiding the design of ideal quantum photonic devices. By combining theoretical rigor with experimental advancements, this work paves the way for scalable, tunable, and robust quantum photonic circuits, essential for future applications in quantum communications, computation, and sensing.


\section{Device Structure}
As shown in Figs.~\ref{fig:Design}(a)-(b), the device consists of a ($170$~nm thick) GaAs p-i-n membrane containing a layer of InAs QDs at its center. The membrane is patterned into a hexagonal-lattice photonic crystal with a period ${a=240}$~nm and a hole radius ${r=0.3a}$, creating a photonic bandgap for TE-polarized light spanning $730$-$1050$~nm. AlGaAs tunneling barriers on either side of the QD layer help to confine charge carriers, whereas the top p-doped and bottom n-doped layers enable Stark tuning of the QD emission wavelength via an applied bias.

By omitting an air hole in the center, we form an H1 cavity. Removing a row of holes on opposite sides of the cavity creates two W1 waveguides, which couple to the cavity modes and guide light towards nanobeam waveguides and outcoupler gratings for efficient off-chip collection. The inner holes of the H1 cavity are reduced to a radius ${r_c = 0.21a}$ and displaced outward by $0.09a$, while the inner holes of the W1 waveguides are reduced to a radius ${r_w = 0.27a}$. These adjustments allow for precise control of the cavity mode frequencies and coupling properties. Figure~\ref{fig:Design}(c) shows a scanning electron microscope (SEM) image of the fabricated structure, overlaid with the simulated time-averaged electric field intensity, illustrating the mode profiles for (i) a $\sigma^{-}$ and (ii) a $\sigma^{+}$ dipole placed in the center of the cavity. Additional details on the sample are provided in Supplementary Information Section S5.


\begin{figure}
    \includegraphics[width = 0.5\textwidth]{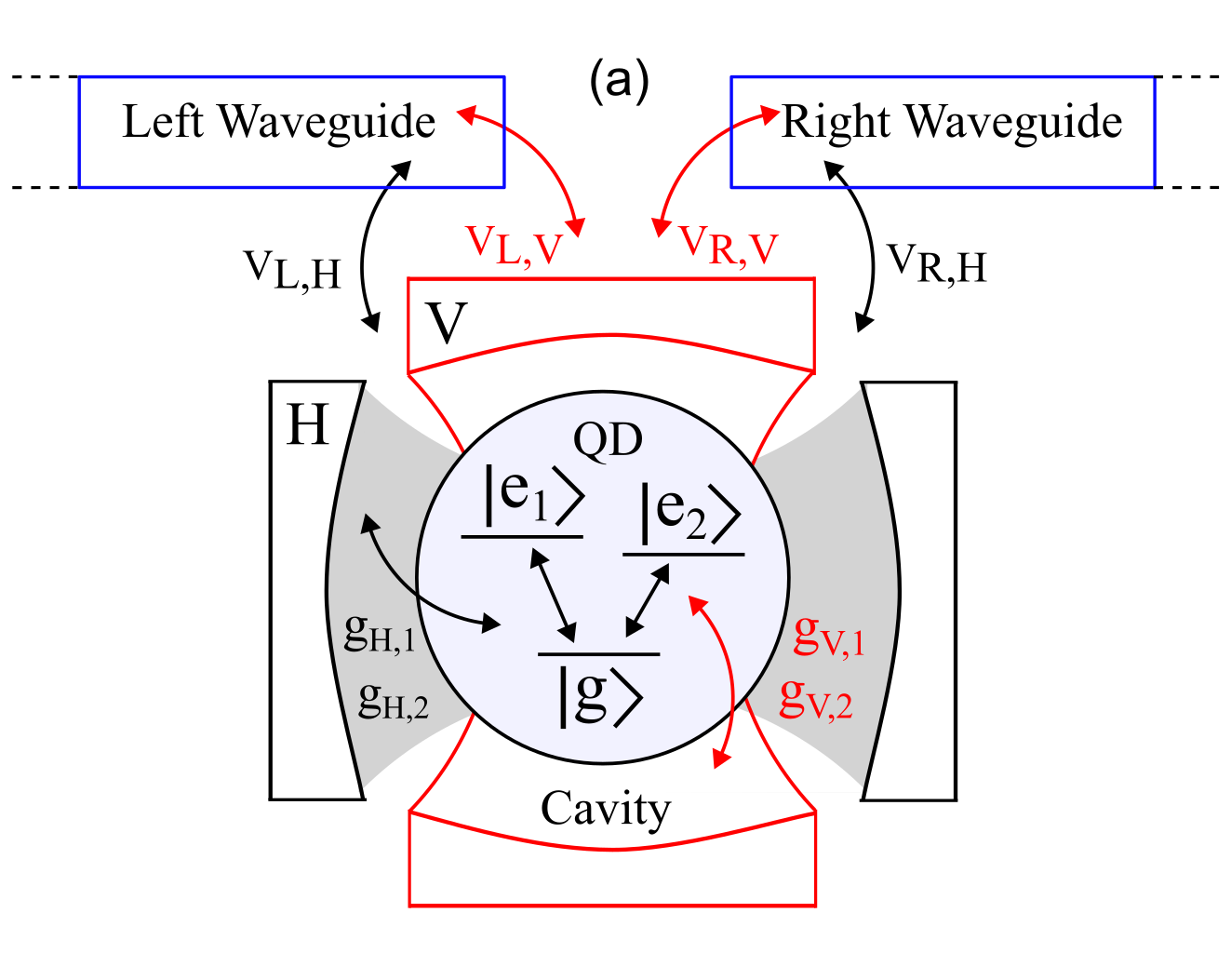}
    \caption{(a) Schematic of the waveguide-coupled nanocavity. Transition $j \in \{1,2\}$ of the QD ($\ket{g} \leftrightarrow \ket{e_j}$) has frequency $\omega_{e,j}$ and couples to cavity mode $\alpha \in \{H,V\}$ (having resonance frequency $\omega_{c,\alpha}$) with coupling rate $g_{\alpha,j}$. Cavity mode $\alpha$ also couples to the right and left waveguides with coupling rates $V_{R,\alpha}$ and $V_{L,\alpha}$, respectively.   }
    \label{fig:theory_diagram}
\end{figure}

\section{Theory \label{sec:theory_trajectories}}
To accurately capture the behavior of the device, our theory must account for realistic QD properties, including fine structure splitting, imperfect crystalline symmetry, and dipole misalignment. These factors lead to deviations from ideal circular polarization and to coherent population transfer between the QD’s excited states. A simple two-level model of a perfectly aligned emitter cannot reproduce the asymmetric directional emission that we observed in our experiments. By including a three-level QD structure, complex coupling rates, and the associated interference effects, our model provides a faithful representation of the QD-cavity-waveguide system. 

In this section, we provide an overview of the theoretical models used to describe the waveguide-coupled nanocavity system and to predict the device's performance. We present the key equations below; more details are provided in the appendices, with detailed derivations given in the Supplementary Information.

As shown in Fig.~\ref{fig:theory_diagram}, we consider a QD with fine structure splitting, modeled as a three-level V-type emitter $\{\ket{g}, \ket{e_1}, \ket{e_2}\}$, coupled to two orthogonal cavity modes $\{H,V\}$, and further coupled to left ($L$) and right ($R$) waveguides. The frequency of the $\ket{g} \leftrightarrow \ket{e_j}$ transition is $\omega_{e,j}$ and the emitter-cavity coupling rates are denoted by $g_{\alpha,j}$, where $j \in \{1,2\}$ labels the QD transitions and $\alpha \in \{H,V\}$ labels the two cavity modes. The mode frequency of the $H$ ($V$) mode is $\omega_{c,H}$ ($\omega_{c,V}$). Furthermore, the cavity-waveguide coupling rates are $V_{\mu,\alpha}$, where $\mu \in \{L,R\}$ labels the left and right waveguides, respectively. The cavity modes are coupled to the two waveguides with equal strength, so we calculate these coupling rates as ${V_{R,\alpha} = V_{L,\alpha} = \omega_{c,\alpha}/2Q_\alpha}$, where $Q_\alpha$ is the quality ($Q$) factor of mode $\alpha$.

The total Hamiltonian for the system takes the form:
\begin{equation}
H = H_e + H_c + H_{\text{wg}} + H_{e\text{-}c} + H_{c\text{-wg}},
\label{eq:H}
\end{equation}
where the emitter Hamiltonian $H_e$, the cavity Hamiltonian $H_c$, the waveguide Hamiltonian $H_{\text{wg}}$, the emitter-cavity interaction $H_{e\text{-}c}$, and the cavity-waveguide interaction $H_{c\text{-wg}}$ are given in Appendix~A. Using the input-output formalism~\cite{Gardiner1985, Fan2010}, we derive the single-photon scattering matrix elements for the waveguide-to-waveguide transmission, e.g.,
\begin{equation}
S_{pk}^{RL} = t_{RL}\,\delta(p-k),
\end{equation}
where $k$ and $p$ denote the input and output photon frequencies, respectively, and $t_{RL}$ is the transmission coefficient for left-to-right transmission (see Appendix B, and Supplementary Information Section S1 for the derivation and result). From this, we obtain the transmission probability $|t_{RL}|^2$, and use it to fit experimental data from our transmission measurements.

To predict the directional contrast of the system, we use a space-discretized waveguide model based on quantum trajectory theory~\cite{Tian1992,Regidor2021,Crowder2022}, which allows us to simulate the time evolution of the system and calculate the probability of a photon being emitted into each of the waveguides. In the model, each waveguide is discretized into a series of $N$ boxes. At each time step in the quantum trajectory algorithm, the cavity interacts with the first box in each waveguide (box $0$), photon number measurements are simulated on the final box in each waveguide (box ${N-1}$), the quantum state of the system is projected according to the measurement results, and the boxes are moved along by one (see Appendix C and Supplementary Information Section S2 for more details). Each trajectory is a stochastic evolution process, so we take an average over many trajectories to compute the expected time evolution of observables (conditioned on photon detection events at the waveguide ends). The contrast $C$ is then given by the normalized difference between the final waveguide box populations at the end of the trajectories (i.e., at time ${t = t_\text{end}}$):
\begin{equation}
C = \left| \frac{ \bigl\langle A_{R,N-1}^{\dag} A_{R,N-1}^{\vphantom{\dag}} \bigr\rangle_{t=t_{\text{end}}} - \bigl\langle A_{L,N-1}^{\dag} A_{L,N-1}^{\vphantom{\dag}} \bigr\rangle_{t=t_{\text{end}}} }{ \bigl\langle A_{R,N-1}^{\dag} A_{R,N-1}^{\vphantom{\dag}} \bigr\rangle_{t=t_{\text{end}}} + \bigl\langle A_{L,N-1}^{\dag} A_{L,N-1}^{\vphantom{\dag}} \bigr\rangle_{t=t_{\text{end}}} } \right|,
\label{eq:theory_contrast}
\end{equation}
where $A_{\mu,N-1}^{\vphantom{\dag}}$ ($A_{\mu,N-1}^{\dag}$) is the annihilation (creation) operator for the final box of waveguide $\mu$ (box ${N-1}$). The end time $t_{\text{end}}$ is chosen such that the emitter has fully decayed to its ground state $\ket{g}$ by the end of the simulations, and all of the photon population has been transferred to the ends of the waveguides. Eq.~(\ref{eq:theory_contrast}) allows us to predict the directional contrast of our system for a given initial state and chosen parameters, and therefore to study how the energy level structure of the QD and the coupling rates affect the degree of directionality.

From our experimental results, we can extract the contrast using the relative power transmitted through the waveguides when exciting the QD in the cavity:
\begin{equation}
C_{\sigma^+/\sigma^-} =\left| \frac{P_{L/R} - P_{R/L}}{P_{L/R} + P_{R/L}}\right|,
\label{contrast}
\end{equation}
where $P_{R}$ ($P_{L}$) is the power transmitted to the right (left). This enables us to quantify the directional performance of our device and to compare the experimental results with theoretical predictions from the quantum trajectory model.


\section{Results}
In this section, we present results for two devices (labeled Device~1 and Device~2), where a demonstration of the device's wavelength-dependent directionality and Purcell enhancement properties is shown.


\subsection{Device 1: Quantum Dot with Simultaneous High Enhancement and Directional Contrast}

\begin{figure}[t!]
\centering
\includegraphics[width=0.5\textwidth]{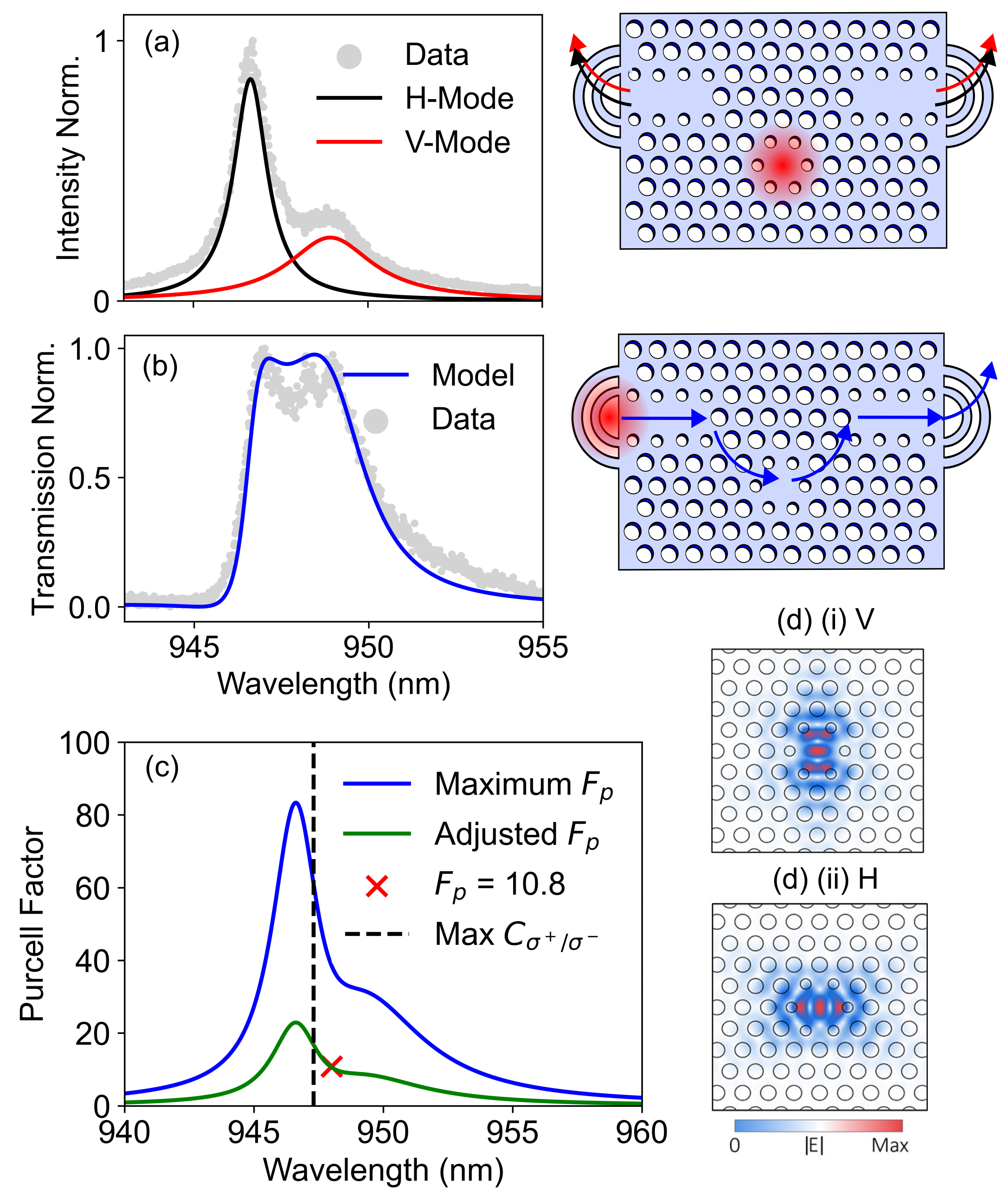}
\caption{\label{fig:DeviceOne_1} (a) Measurement of the cavity modes of Device~1, with the fits to the data being two Lorentzian peaks. (b) Measured waveguide-to-waveguide transmission through Device~1, fit using our scattering matrix model. (c) Purcell factor as a function of wavelength for Device~1. The blue curve represents the calculated maximum Purcell factor $F_p$, while the green curve shows the adjusted Purcell factor accounting for the QD's offset position and non resonant excitation, calculated from the experientially measured result. The red cross highlights the measured Purcell factor of ${F_p = 10.8}$ at $948$~nm. The dashed vertical line at $947.3$~nm indicates the wavelength where the directional contrast is predicted to peak at $98$\%, with the corresponding Purcell enhancement estimated to be $16.8$. (d) The electric field profiles of the (i) $V$ and (ii) $H$ cavity modes. Regions of higher electric field concentration correspond to a greater Purcell enhancement.}
\end{figure}

\begin{figure*}[t!]
\centering
\includegraphics[width=0.99\textwidth]{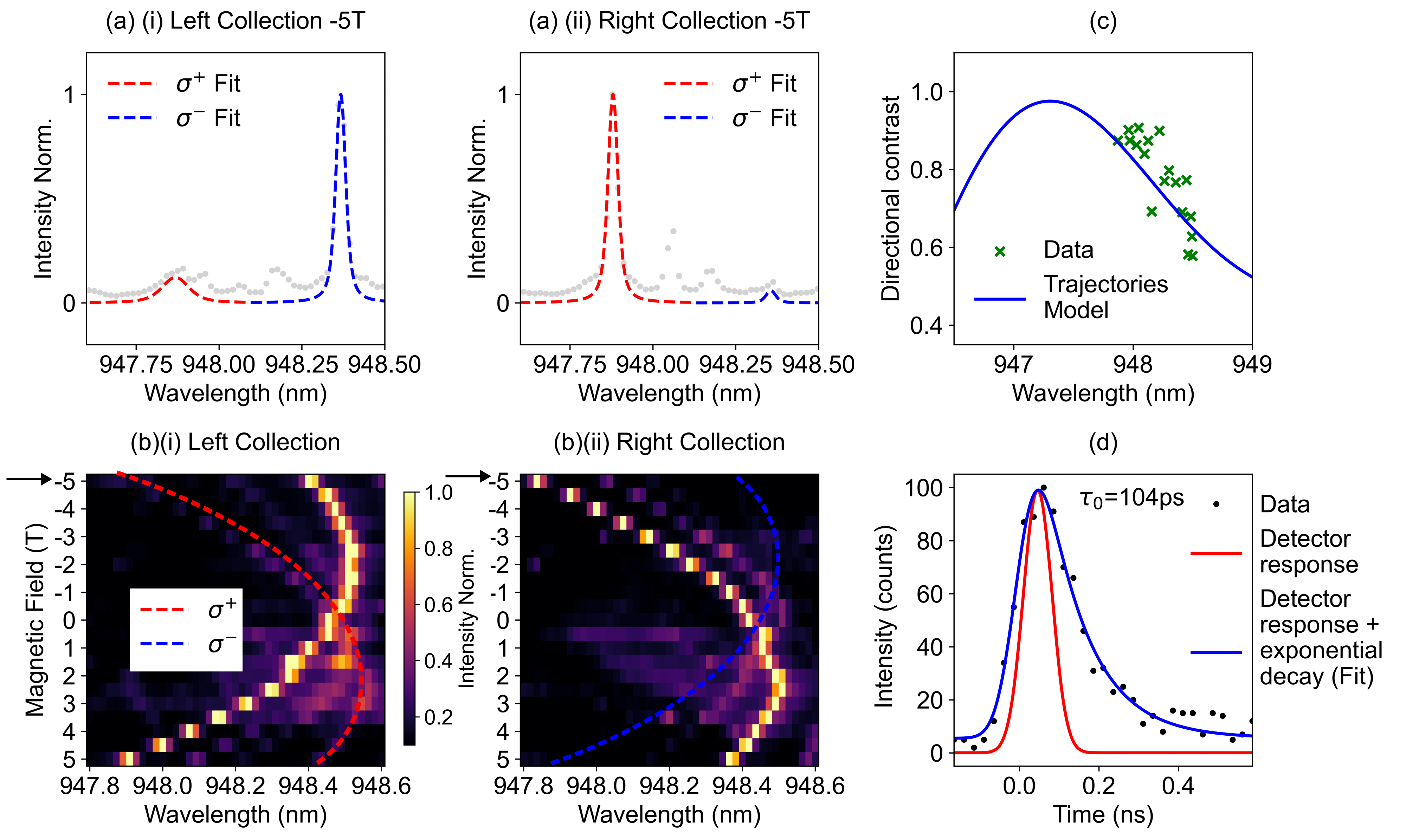}
\caption{\label{fig:DeviceOne} (a) PL spectra, exciting the cavity, collected from the (i) left and (ii) right outcouplers, with an applied magnetic field  $B = -5$T. The Zeeman-split emission lines of the QD are fit using Voigt functions. (b) Magnetic field dependence of PL emission from the QD, measured from the (i) left and (ii) right outcouplers. The dashed lines mark the wavelengths of the two QD spin states. (c) Directional contrast of a charged exciton state of the QD in Device~1, as a function of wavelength. The QD emission wavelength was tuned through the application of a magnetic field from $-5$T to $5$T in a Faraday geometry. The prediction of the directional contrast for a circularly polarized emitter from the quantum trajectory model (blue curve) is presented in comparison to the measured data (green crosses). (d) Measurement of the QD decay time. The instrument response function is indicated by a red line, and the blue fit is an exponential decay convolved with the instrument response.}
\end{figure*}

Achieving both large Purcell enhancement and highly directional emission in solid-state QEs has been a longstanding challenge. Here, Device~1 overcomes this, demonstrating a directional contrast of ${C = 88\pm1\%}$ and a Purcell factor of ${F_P = 10.8\pm0.7}$. 

The mode properties of the cavity are measured by exciting the cavity using a $808$~nm laser and collecting photoluminescence (PL) from one outcoupler. The intensity of PL as a function of wavelength for Device~1 is shown in Fig.~\ref{fig:DeviceOne_1}(a). By fitting the data with a Lorentzian function, we extract a resonant wavelength of $946.63$~nm ($948.93$~nm) and a $Q$ factor of $810$ ($320$) for the $H$ ($V$) mode, respectively. From these $Q$ factors and resonant wavelengths, we are able to predict the maximum directional contrast possible for a circularly polarized emitter at the cavity center using the trajectories model, as well as the expected transmission from the scattering matrix model, as shown in Fig.~\ref{fig:DeviceOne}(c) and Fig.~\ref{fig:DeviceOne_1}(b) respectively. The transmission data presented in Fig.~\ref{fig:DeviceOne_1}(b) is obtained by exciting one outcoupler of the device with a laser, to excite a broad QD ensemble, and measuring the PL intensity at the opposite outcoupler. This experimental result is compared with the expected transmission for a device with these cavity mode properties, as predicted by our transmission model. Figure~\ref{fig:DeviceOne_1}(c) shows the expected maximum Purcell factor in the center of the cavity for a circularly polarized emitter, modeled based on the mode data presented in Fig.~\ref{fig:DeviceOne_1}(a) (see Appendix~D). Figure~\ref{fig:DeviceOne_1}(d) illustrates the electric field profiles of cavity modes (i) $V$ and (ii) $H$, where regions of higher electric field concentration correspond to a greater Purcell enhancement, with the maximum enhancement occurring in the center of the cavity.

\begin{figure*}[t]
\centering
\includegraphics[width=0.95\textwidth]{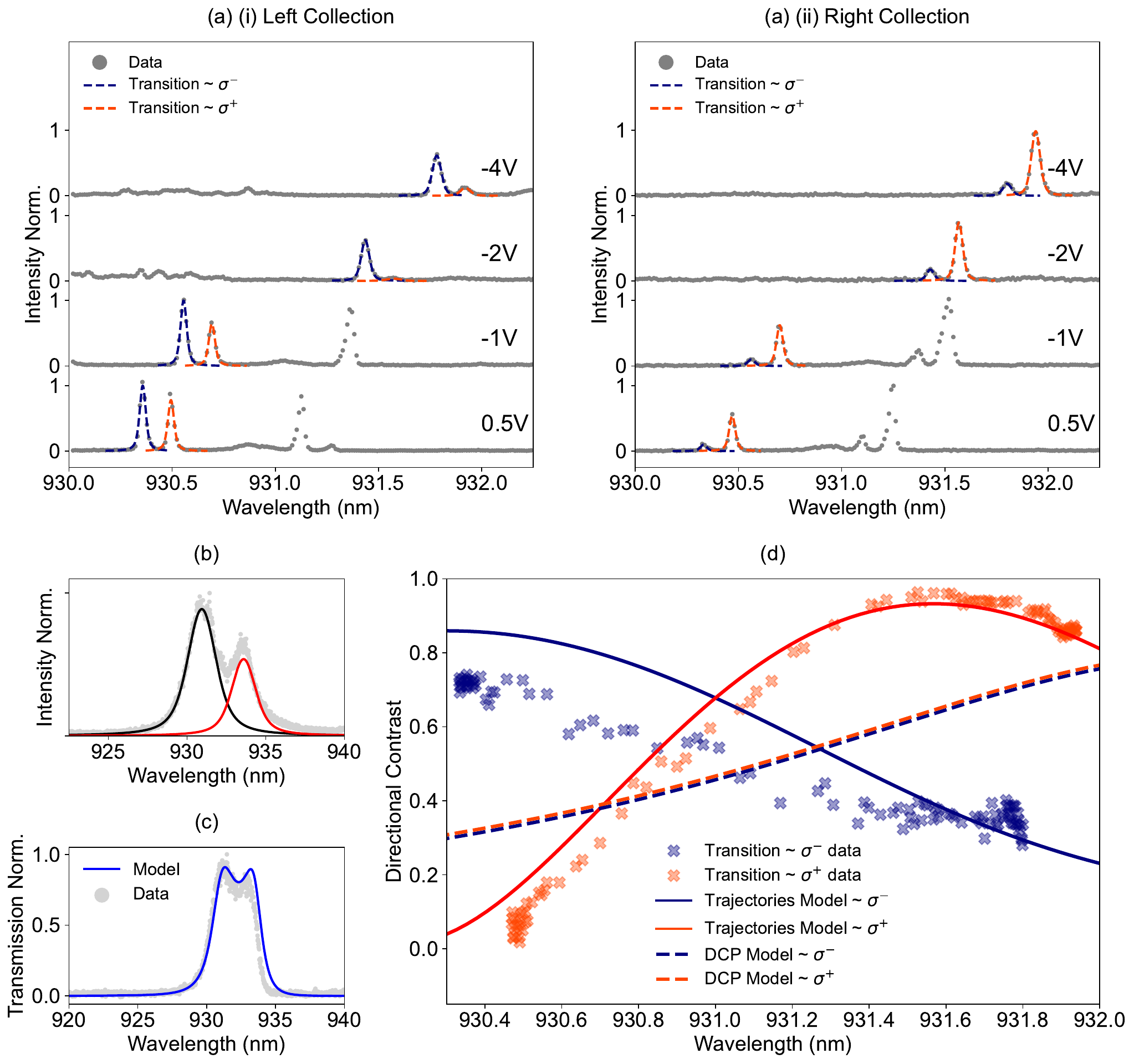}
\caption{\label{fig:DeviceTwo} (a) Individual PL spectra at $-4$V, $-2$V, $-1$V, and $0.5$V for collection from the (i) left and (ii) right outcoupler, showing the individual QD emission lines. The QD lines are fit using a Voigt function. (b) Measurement of the H1 cavity modes of Device~2; the data is fit by two Lorentzian peaks. (c) Measured waveguide-to-waveguide transmission through Device~2, fit using our transmission model. (d) Directional contrast for Device~2 as a function of emitter wavelength. The contrast is measured for the Zeeman-split lines of a QD within a $1.5$T $B$-field. The data is compared with predictions from the quantum trajectory model (solid curves), as well as with the DCP at the cavity center (dashed curves), which is the contrast expected for a circularly polarized emitter~\cite{ChiralCav_2022}.}
\end{figure*}

Figure~\ref{fig:DeviceOne}(b)(i-ii) shows the magnetic field dependence (ranging from $-5$T to $5$T) of PL emission from the QD in Device~1, collected from the two outcouplers. The directional behavior of this emission is clearly visible across the magnetic field range, as the applied magnetic field creates well-resolved $\sigma^{+}$ and $\sigma^{-}$ dipoles. Figure~\ref{fig:DeviceOne}(a)(i-ii) shows the individual spectra for $-5$T, where it can clearly be seen that the emission of the $\sigma^{+}$ state is predominantly to the right outcoupler, while the emission of the $\sigma^{-}$ state couples predominantly to the left outcoupler. In Fig.~\ref{fig:DeviceOne}(c), the experimentally measured directional contrast is plotted as a function of the QD emission wavelength (green crosses), which was controlled by tuning the applied magnetic field strength. The magnetic field is capable of tuning the QD wavelength over a range of $0.63$~nm, from $947.87$~nm to $948.5$~nm. In the figure, the measured directional contrast is compared to what is expected for a device with the measured mode properties (blue curve). We see that the measured lower contrast at longer wavelengths matches our trajectories model for the wavelength dependence of the contrast. 

Figure~\ref{fig:DeviceOne}(d) shows a measurement of the QD decay time. In this measurement, the QD is excited with a fs laser pulse at $810$~nm. The QD transition at $948.0$~nm exhibits a decay time ${\tau_0 = (104\pm 2})$~ps within an applied magnetic field of ${B=2}$T. The instrument response function is a Gaussian with a full width at half-maximum of $50$~ps. From measurements of the decay time for $10$ QDs in bulk GaAs in this sample, we extract an average ensemble lifetime $\tau_{\text{bulk}} = (1.13 \pm 0.08)$~ns, and therefore estimate the Purcell factor of the QD in the cavity to be ${F_P = \tau_{\text{bulk}}/\tau_0 = 10.8\pm0.7}$. We note that the QD lifetime was measured using above-band excitation, which introduces a contribution to the decay time from inter-band processes. Since the QD was not visible under resonant excitation, we were unable to directly measure the resonant decay time. Therefore, the measured value represents only a lower bound for the true Purcell factor. Moreover, in this device, it was not possible to tune the QD to shorter wavelengths. At $947.3$~nm, where the directional contrast is predicted to peak at $98$\%, the Purcell enhancement using an above-band excitation scheme is predicted to be $16.8$, as determined from Eq.~(\ref{purcell}) in Appendix D.


\subsection{Device 2: Quantum Dot with Highly Tunable Asymmetric Directional Contrast}

Figure~\ref{fig:DeviceTwo} summarizes the experimental results from Device~2. In this device, the QDs exhibited longer decay times and could be tuned over a wavelength range of approximately $1.75$~nm, making them ideal for studying and controlling the directional contrast. Figure~\ref{fig:DeviceTwo}(b) displays the cavity modes, with resonant wavelengths of $930.9$~nm for the $H$ mode and $933.6$~nm for the $V$ mode, and corresponding $Q$ factors of $660$ and $779$. The measured transmission through the device, shown in Fig.~\ref{fig:DeviceTwo}(c), agrees well with the predictions of our transmission model.

Figure~\ref{fig:DeviceTwo}(a)(i-ii) shows the PL spectra of a QD transition collected from the left and right outcouplers under varying applied voltages and a $1.5$~T magnetic field. The magnetic field induces Zeeman splitting, resulting in two distinct components labeled as $\sigma^{+}$ and $\sigma^{-}$. The PL spectra show how the emission intensities for each Zeeman component shift as the applied voltage varies between $-4$V and $0.5$V. In Fig.~\ref{fig:DeviceTwo}(a)(i), the normalized PL intensities collected from the left outcoupler reveal that the relative strength of the $\sigma^{+}$ and $\sigma^{-}$ transitions depends strongly on the applied bias. Similarly, Fig.~\ref{fig:DeviceTwo}(a)(ii) demonstrates this voltage-dependent behavior in the PL spectra collected from the right outcoupler, with complementary changes in intensity.

By normalizing the PL spectra from the left and right outcouplers and applying Eq.~(\ref{contrast}), we compute the wavelength-dependent directional contrast, which is shown in Fig.~\ref{fig:DeviceTwo}(d). For the emission line corresponding to $\sigma^{+}$, the directional contrast can be dynamically tuned from $2\%$ to $96\%$. This significant tunability is achieved through the quantum-confined Stark effect, where the applied bias alters the QD's energy levels, shifting the emission wavelength and modulating the contrast. The ability to electrically tune the directional contrast provides precise control over the optical properties of the device, making it highly suitable for dynamic photonic applications such as tunable optical isolators or directional emitters.

However, the observed directional contrast in Fig.~\ref{fig:DeviceTwo}(d) deviates significantly from the predictions of the contrast based solely on the degree of circular polarization (DCP) of the electric field in the cavity~\cite{ChiralCav_2022} (dashed curves). A striking feature in Device~2 is the asymmetry in contrast between the Zeeman components of the QD states, an effect that finite-difference time-domain (FDTD) simulations and a simple DCP analysis do not predict. This clear discrepancy underscores the limitations of the DCP model in capturing complex emitter-cavity interactions and directional asymmetries. Similar asymmetries have been reported in previous studies~\cite{NB1_2016,Topological2_2020,Topological4_2020,add_drop,Javadi2018}.

To resolve this, our quantum trajectories model, unlike the DCP approach, incorporates fine-structure splitting and accounts for variations in emitter-cavity coupling rates. By including these factors, the model successfully reproduces the experimentally observed asymmetry, demonstrating its predictive capability, and highlighting its ability to capture the complex physics behind wavelength-dependent directional contrast.

As shown in Fig.~\ref{fig:DeviceOne_1}(d), the cavity modes $H$ and $V$ exhibit distinct regions of electric field concentration. Near the center of the cavity, the emitter experiences symmetric coupling to both modes, minimizing mismatch. However, displacements greater than 60\,nm from the center introduce significant spatial mismatches, thus altering the Purcell factors for the $H$ and $V$ modes. Another consideration is that misalignment of the emitter from the crystal axis will alter its intrinsic dipole orientation relative to the cavity field, causing the emitted photons to acquire mixed or elliptical polarization states rather than pure circular \(\sigma^+/\sigma^-\) polarizations. This mixing changes how the emitter couples to the cavity modes and can affect both the coupling efficiencies and the resulting directional emission characteristics. Additionally, deviations from ideal dot symmetry can further modify the polarization states and coupling rates. In Fig.~\ref{fig:DeviceTwo}(d), the solid orange and blue lines represent the contrast predicted by our quantum trajectories model, aligning closely with the experimental asymmetry observed in Fig.~\ref{fig:DeviceTwo}(a)(i-ii). A more detailed discussion of these dynamics and the coupling rate parameters used in this figure are provided in Supplementary Information Sections S3 and S6.


\section{Conclusion}

In conclusion, our work has successfully demonstrated a waveguide-coupled cavity nanophotonic device that achieves a remarkable balance between near-unity directional coupling and significant Purcell enhancement for an embedded circularly polarized emitter. The low mode volume of the H1 cavity, approximately \({V \approx 0.6(\lambda/n)^{3}}\)~\cite{H1Cav1_2015}, is critical in achieving a large emitter-cavity coupling strength within a relatively low-\(Q\) (\({Q \approx 400\text{-}1000}\)) cavity environment. The performance of the device is robust against both spectral and spatial perturbations of the embedded emitter, a significant advance over alternative approaches~\cite{ChiralCav_2022}. In particular, as shown in Ref ~\cite{ChiralCav_2022}  the directional contrast of our device remains exceptionally high even with emitter displacements up to $60$~nm from the center of the cavity. This feature underlines the potential for the device in practical applications where precise emitter positioning can be challenging.

Furthermore, our experiments on two devices reveal that the directional contrast strongly depends on the wavelengths of the cavity modes and the emitter. This sensitivity allows post-fabrication tuning and optimization \cite{StrainTuning}, demonstrating the device's adaptability to varying emitter properties while maintaining high performance. We present results showing precise tuning of a QD emission line from a directional contrast of \(2\%\) to \(96\%\). Notably, our quantum trajectory model explains for the first time the observed asymmetry in directional contrast seen in QD-waveguide devices and provides a clear pathway for optimizing QD-based directional light–matter interfaces, even in the face of imperfect polarization.

Lastly, the general mechanism behind our directional light–matter interaction is not confined to our specific cavity design. It extends to other cavity architectures that support degenerate orthogonal modes, such as micropillars~\cite{Wang2019}, and would be effective for systems operating in the telecom bands~\cite{Phillips2024,Kim2016}. This broad applicability could pave the way for a range of high-performance applications in quantum nanophotonics, including reconfigurable phase shifters~\cite{Reconfig}, quantum routers and switches~\cite{Kimble2008}, and compact optical circulators~\cite{doi:10.1126/science.aaj2118}.


\section*{Data availability}
Data underlying the results presented in this paper are not publicly available at this time but can be obtained from the authors on reasonable request.

\section*{Acknowledgments}
The authors acknowledge helpful discussions with Mahmoud Jalali Mehrabad and Yuxin Wang. This work was supported by EPSRC Grant No. EP/N031776/1, EP/W524360/1, EP/R513313/1 and EP/V026496/1.

\section*{Author contributions statement}
D.H. designed the photonic structures, which R.D. fabricated.  N.J.M., D.H., L.H., L.B. and M.D. carried out the measurements and simulations. E.C. and M.D. developed the scattering matrix model for transmission. M.D. derived the quantum trajectory model. L.R.W., P.K. and M.S.S. provided supervision and expertise. N.J.M., D.H., L.B. and M.D. wrote the manuscript, with input from all authors. N.J.M. and D.H. contributed equally to this work.


\section*{Appendix}
In this Appendix, we present details of our theoretical models that are relevant to the discussions in the main text. More detailed calculations are contained in the Supplementary Information.


\subsection*{Appendix A: System Hamiltonian}

We consider the system depicted in Fig.~\ref{fig:theory_diagram} in the main text. In Eq.~(\ref{eq:H}), we have the free emitter Hamiltonian (${\hbar=1}$)
\begin{equation}
H_e = \sum_{j=1,2} \omega_{e,j} \sigma_j^+ \sigma_j^-
\end{equation}
(where $\sigma_j^+ = \ketbra{e_j}{g}$ and $\sigma_j^- = \ketbra{g}{e_j}$ are the raising and lowering operators for transition ${j \in \{1,2\}}$), the free cavity Hamiltonian
\begin{equation}
H_c = \sum_{\alpha=H,V} \omega_{c,\alpha} c_{\alpha}^{\dag} c_{\alpha}^{\vphantom{\dag}}
\end{equation}
(where $c_{\alpha}$ and $c_{\alpha}^{\dag}$ are the annihilation and creation operators for mode ${\alpha \in \{H,V\}}$), and the free waveguide Hamiltonian
\begin{equation}
H_{\text{wg}} = \sum_{\mu = L,R} \int \omega(k) a_{\mu}^{\dag}(k) a_{\mu}^{\vphantom{\dag}}(k) dk,
\end{equation}
where the operators $a_{\mu}(k)$ and $a_{\mu}^{\dag}(k)$ annihilate and create photons with wave number $k$ in waveguide ${\mu \in \{L,R\}}$. This Hamiltonian assumes identical dispersion relations $\omega(k)$ for both waveguides. 

The emitter-cavity interaction is of the Jaynes-Cummings form:
\begin{equation}
H_{e\text{-}c} = \sum_{j=1,2} \sum_{\alpha = H,V} \left( g_{\alpha,j}^{\vphantom{*}} \sigma_j^+ c_{\alpha}^{\vphantom{\dag}} + g_{\alpha,j}^* \sigma_j^- c_{\alpha}^{\dag} \right),
\end{equation}
and the cavity-waveguide interaction, under the Markov approximation~\cite{Gardiner1985} (assuming coupling independent of $k$), is:
\begin{align}
\begin{split}
H_{c\text{-wg}} = \sum_{\alpha = H,V} \sum_{\mu = L,R} &\int \biggl[ \sqrt{\frac{V_{\mu,\alpha}}{2\pi}} a_{\mu}^{\dag}(k) c_{\alpha}^{\vphantom{\dag}} e^{i\pi \delta_{\alpha H}\delta_{\mu L}} \\
&+ \sqrt{\frac{V_{\mu,\alpha}^*}{2\pi}} a_{\mu}^{\vphantom{\dag}}(k) c_{\alpha}^{\dag} e^{-i\pi \delta_{\alpha H}\delta_{\mu L}} \biggr] dk.
\end{split}
\end{align}
The phase factors $e^{\pm i\pi \delta_{\alpha H} \delta_{\mu L}}$ reflect the anti-symmetric coupling of the cavity modes into the left and right waveguides, which gives rise to the interference that enables directional emission in our device~\cite{ChiralCav_2022}.


\subsection*{Appendix B: Scattering Matrix Model}

The single-photon scattering matrix elements are given by:
\begin{equation}
S_{pk}^{\mu\nu} = \langle p_{\mu}|S|k_{\nu}\rangle = \braket{p_{\mu}^-}{k_{\nu}^+} = \expval{a_{\mu,\text{out}}(p) a_{\nu,\text{in}}^{\dag}(k)}{0},
\end{equation}
where $k$ is the frequency of the input photon in waveguide $\nu$, $p$ is the frequency of the output photon in waveguide $\mu$, ${\ket{k_{\nu}^+} = a_{\nu,\text{in}}^{\dag}(k)\ket{0}}$ and ${\ket{p_{\mu}^-} = a_{\mu,\text{out}}^{\dag}(p)\ket{0}}$ are scattering eigenstates, and $\ket{0}$ is the vacuum state of the system. For transmission, we set ${\mu \neq \nu}$, such that the output photon leaves in the opposite waveguide to the input photon. For example, for left-to-right transmission, ${\nu = L}$ and ${\mu = R}$, and the corresponding transmission matrix element has the form:
\begin{equation}
S_{pk}^{RL} = \expval{a_{R,\text{out}}(p) a_{L,\text{in}}^{\dag}(k)}{0} = t_{RL}\,\delta(p-k),
\end{equation}
where $t_{RL}$ is the transmission coefficient, and ${\delta(p-k)}$ enforces energy conservation. The derivation of the scattering matrix is presented in Supplementary Information Section S1, where we obtain the transmission coefficient for both left-to-right and right-to-left propagation, and demonstrate that the directional light–matter interaction in our cavity can give rise to nonreciprocal phase shifts.


\subsection*{Appendix C: Quantum Trajectory Model}

In the quantum trajectory model, each waveguide is discretized into a series of $N$ spatial bins of width $\Delta t$ in time (labeled with the index ${n \in \{0,1,\dotsc,N-1\}}$), allowing the time evolution to be solved numerically in discrete time steps. In the Supplementary Information (Section S2), we discretize the Hamiltonian from Eq.~(\ref{eq:H}) (with the terms defined in Appendix A), where we replace the continuum waveguide mode operators $\smash{a_{\mu}^{\vphantom{\dag}}(k)}$, $\smash{a_{\mu}^{\dag}(k)}$ with discrete mode operators $\smash{a_{\mu,k}^{\vphantom{\dag}}}$, $\smash{a_{\mu,k}^{\dag}}$, and use the discrete Fourier transforms
\begin{equation}
A_{R,n} = \frac{1}{\sqrt{N}} \sum_{k=0}^{N-1} a_{R,k} e^{i \omega_k n \Delta t}
\label{eq:FT_R}
\end{equation}
and
\begin{equation}
A_{L,n} = \frac{1}{\sqrt{N}} \sum_{k=0}^{N-1} a_{L,k} e^{-i \omega_k n \Delta t}
\label{eq:FT_L}
\end{equation}
to transform from $k$-space to position space ($\omega_k$ is the discretized waveguide dispersion relation). The operator $A_{\mu,n}^{\vphantom{\dag}}$ ($A_{\mu,n}^{\dag}$) annihilates (creates) a photon in spatial bin $n$ of waveguide $\mu$. After discretizing the Hamiltonian, we express all operators as matrices by choosing the following basis consisting of states with at most one photon in the system (since we consider single-photon emission from the cavity):
\begin{align}
\begin{split}
\Bigl\{ &\ket{g, 0_H, 0_V, 0_{R,N-1}, \dotsc, 0_{R,0}, 0_{L,N-1}, \dotsc, 0_{L,0}}, \\
   &\ket{e_1, 0_H, 0_V, 0_{R,N-1}, \dotsc, 0_{R,0}, 0_{L,N-1}, \dotsc, 0_{L,0}}, \\[0.05in]
   &\ket{e_2, 0_H, 0_V, 0_{R,N-1}, \dotsc, 0_{R,0}, 0_{L,N-1}, \dotsc, 0_{L,0}}, \\[0.05in]
   &\ket{g, 1_H, 0_V, 0_{R,N-1}, \dotsc, 0_{R,0}, 0_{L,N-1}, \dotsc, 0_{L,0}}, \\[0.05in]
   &\ket{g, 0_H, 1_V, 0_{R,N-1}, \dotsc, 0_{R,0}, 0_{L,N-1}, \dotsc, 0_{L,0}}, \\[0.05in]
   &\ket{g, 0_H, 0_V, 1_{R,N-1}, \dotsc, 0_{R,0}, 0_{L,N-1}, \dotsc, 0_{L,0}}, \\
   &\hspace{1.3in} \vdots\\
   &\ket{g, 0_H, 0_V, 0_{R,N-1}, \dotsc, 0_{R,0}, 0_{L,N-1}, \dotsc, 1_{L,0}} \Bigr\}.
\label{eq:basis}
\end{split}
\end{align}
The basis states are tensor products of states from the emitter basis, the cavity basis, and the discrete position-space waveguide basis, where $\ket{0_\alpha}$ ($\ket{1_\alpha}$) corresponds to no photons (one photon) in cavity mode $\alpha$, and $\ket{0_{\mu,n}}$ ($\ket{1_{\mu,n}}$) corresponds to no photons (one photon) in spatial bin $n$ in waveguide $\mu$.

At the start of each trajectory simulation, we set the emitter to be in a superposition of its excited states $\ket{e_1}$ and $\ket{e_2}$, and observe how the population of the excited states decays into the waveguides as the system evolves in time. The initial state $\ket{\psi(0)}$ is therefore written in our basis as:
\begin{align}
\begin{split}
\ket{\psi(0)} &= \alpha \ket{e_1,0_H,0_V,0_{R,N-1},\dots,0_{L,0}} \\
&\quad + \beta \ket{e_2,0_H,0_V,0_{R,N-1},\dots,0_{L,0}},
\end{split}
\end{align}
with the normalization condition ${|\alpha|^2 + |\beta|^2 = 1}$. To model the charged exciton QD from Device~1, we consider a two-level system with initial condition ${\alpha = 1}$, ${\beta = 0}$. For the neutral exciton states of Device~2, we consider a three-level system (with fine structure splitting) and initial condition ${\alpha = \beta = 1/\sqrt{2}}$, where we assume that the two excited states are populated equally during non-resonant excitation.

At each time step in the quantum trajectory algorithm, the cavity interacts with the first box (${n=0}$) in each waveguide, photon number measurements are simulated on the final waveguide boxes (${n=N-1}$), and the boxes are moved along by one (see Supplementary Information for more details). In order to obtain the directional contrast $C$ in Eq.~(\ref{eq:theory_contrast}) in the main text, we calculate the expectation values
\begin{equation}
\langle A_{\mu,N-1}^{\dag} A_{\mu,N-1}^{\vphantom{\dag}} \rangle = \bra{\psi(t)} A_{\mu,N-1}^{\dag} A_{\mu,N-1}^{\vphantom{\dag}} \ket{\psi(t)}
\end{equation}
for ${\mu \in \{L,R\}}$ at each time step, which are the populations of the final waveguide boxes. From these, we calculate $C$ at the end of the trajectories (${t = t_{\text{end}}}$), once the emitter has fully decayed and all the photon population has reached the ends of the waveguides.


\vspace{0.1in}
\subsection*{Appendix D: Purcell Enhancement}

Considering the two modes of our cavity system, a circularly polarized emitter will emit into the modes as a proportion of the Purcell factor of that mode (${\alpha \in \{H,V\}}$):
\begin{equation}
    f_{\alpha}=\frac{F_{\alpha}}{F_{H}+F_{V}}.
    \label{purcell}
\end{equation}
The total Purcell factor for a circularly polarized emitter in the cavity is given by:
\begin{equation}
    F_{\text{tot}}=\frac{F_{H}^2+F_{V}^2}{F_{H}+F_{V}}, 
\end{equation}
\noindent
where, for ${\alpha \in \{H,V\}}$,
\begin{equation}
F_{\alpha}=\frac{{3Q_{\alpha}}}{{4{\mathrm{\pi}}^{2}V_{\alpha}}}\frac{{(2\kappa_{\alpha} )^2}}{{4\left({\omega_e - \omega_{c,\alpha} }\right)^2 + (2\kappa_{\alpha} )^2}}\frac{\left|{\bm{\mu}} \cdot {\mathbf{E}}({\mathrm{r}_0}) \right|^2}{\left|{\bm{\mu}}\right|^2\left|{{\mathbf{E}}_{{\mathrm{max}}}}\right|^2}.
\end{equation}
Here, $Q_{\alpha}$ is the $Q$ factor of cavity mode $\alpha$, and $V_{\alpha}$ is the mode volume in cubic wavelengths. In addition, $\omega_e$ is the emitter transition frequency, $\omega_{c,\alpha}$ is the mode frequency of cavity mode $\alpha$, and $2\kappa_{\alpha}$ is the full width at half-maximum of the cavity mode. Furthermore, $\bm{\mu}$, ${\mathbf{E}}({\mathrm{r}_0})$, and ${\mathbf{E}}_{{\mathrm{max}}}$ are the transition dipole moment, electric field at the emitter's position, and the maximum electric field in the cavity, respectively.

\section*{Supplemental document}

See Supplement 1 for supporting content.

\bibliography{Bibli}

\widetext
\clearpage

\begin{center}
\textbf{\large Supplementary Information: Purcell-Enhanced, Directional Light–Matter Interaction in a Waveguide-Coupled Nanocavity}
\end{center}
\newpage
\setcounter{equation}{0}
\setcounter{figure}{0}
\setcounter{table}{0}
\setcounter{page}{1}
\setcounter{section}{0}
\makeatletter
\renewcommand{\theequation}{S\arabic{equation}}
\renewcommand{\thefigure}{S\arabic{figure}}

\renewcommand{\thesection}{S\arabic{section}}



\newpage
\section{Scattering Matrix Model for Transmission}

In this section of the Supplementary Information, we present the derivation of the scattering matrix model that we use to calculate the waveguide-to-waveguide transmission in our device. In Section~\ref{subsec:transmission_Hamiltonian}, we transform the Hamiltonian given in the main text from $k$-space to frequency space, which then allows us to derive the input-output relations for our system in Section~\ref{subsec:transmission_input_output}. In Section~\ref{subsec:transmission_scattering_matrix}, we use the input-output relations to derive the single-photon scattering matrix for the waveguide-coupled cavity, from which we obtain the transmission. We then compare the transmission result to finite-difference time-domain (FDTD) simulations in Section~\ref{subsec:transmission_simulations}, and present some results for resonant photon-emitter scattering in Section~\ref{subsec:transmission_resonant}, which demonstrate the possibility of nonreciprocal $\pi$ phase shifts in our device.


\subsection{Hamiltonian Transformation}\label{subsec:transmission_Hamiltonian}

We first transform the Hamiltonian in Eq.~(1) in the main text from $k$-space to frequency space. For convenience, we write the full $k$-space Hamiltonian here, using the definitions in Appendix A:
\begin{align}
\begin{split}
H =& \sum_{j=1,2} \omega_{e,j}^{\vphantom{\dag}} \sigma_j^+ \sigma_j^- + \sum_{\alpha=H,V} \omega_{c,\alpha}^{\vphantom{\dag}} c_{\alpha}^{\dag} c_{\alpha}^{\vphantom{\dag}} + \sum_{\mu = L,R} \int \omega(k) a_{\mu}^{\dag}(k) a_{\mu}^{\vphantom{\dag}}(k) dk + \sum_{j=1,2} \sum_{\alpha = H,V} \Bigl( g_{\alpha,j}^{\vphantom{*}} \sigma_j^+ c_{\alpha}^{\vphantom{\dag}} + g_{\alpha,j}^* \sigma_j^- c_{\alpha}^\dag \Bigr)\\[0.05in]
&+ \sum_{\alpha = H,V} \sum_{\mu = L,R} \int \Biggl[ \sqrt{\frac{V_{\mu,\alpha}^{\vphantom{*}}}{2\pi}} a_{\mu}^{\dag}(k) c_{\alpha}^{\vphantom{\dag}} e^{i\pi \delta_{\alpha H} \delta_{\mu L}} + \sqrt{\frac{V_{\mu,\alpha}^*}{2\pi}} a_{\mu}^{\vphantom{\dag}}(k) c_{\alpha}^{\dag} e^{-i\pi \delta_{\alpha H} \delta_{\mu L}} \Biggr] dk.
\label{eq:H_k}
\end{split}
\end{align}
In order to transform the Hamiltonian $H$ to frequency space, we use the linear dispersion approximation, where the waveguide dispersion relation $\omega(k)$ is assumed to be linear near some frequency ${\omega_0 = \omega(k_0) = \omega(-k_0)}$. Using the Taylor series expansion of $\omega(k)$ at ${k = \pm k_0}$, we have:
\begin{equation}
\omega(k) \approx \left\{ \begin{array}{lr}
\omega_0 - v_g(k + k_0),\hspace{0.05in} k \approx -k_0, \\[0.05in]
\omega_0 + v_g(k - k_0),\hspace{0.05in}  k \approx k_0, \end{array}\right.
\end{equation}
where $\smash{v_g = \frac{d\omega}{dk}\bigr|_{k_0}}$ is the group velocity. Accounting for the fact that the right waveguide operators $\smash{a_R^{\vphantom{\dag}}(k)}$, $\smash{a_R^{\dag}(k)}$ are defined for ${k<0}$ and the left waveguide operators $\smash{a_L^{\vphantom{\dag}}(k)}$, $\smash{a_L^{\dag}(k)}$ are defined for ${k>0}$ (which arises from the mapping of a single infinite waveguide with right- and left-propagating modes onto two semi-infinite waveguides), we obtain:
\begin{align}
\begin{split}
H =& \sum_{j=1,2} \omega_{e,j}^{\vphantom{\dag}} \sigma_j^+ \sigma_j^- + \sum_{\alpha=H,V} \omega_{c,\alpha}^{\vphantom{\dag}} c_{\alpha}^{\dag} c_{\alpha}^{\vphantom{\dag}} + \int \Bigl[\omega_0 - v_g(k + k_0)\Bigr] a_R^{\dag}(k) a_R^{\vphantom{\dag}}(k) dk + \int \Bigl[\omega_0 + v_g(k - k_0)\Bigr] a_L^{\dag}(k) a_L^{\vphantom{\dag}}(k) dk\\[0.05in]
&+ \sum_{j=1,2} \sum_{\alpha = H,V} \Bigl( g_{\alpha,j}^{\vphantom{*}} \sigma_j^+ c_{\alpha}^{\vphantom{\dag}} + g_{\alpha,j}^* \sigma_j^- c_{\alpha}^\dag \Bigr) + \sum_{\alpha = H,V} \sum_{\mu = L,R} \int \Biggl[ \sqrt{\frac{V_{\mu,\alpha}^{\vphantom{*}}}{2\pi}} a_{\mu}^{\dag}(k) c_{\alpha}^{\vphantom{\dag}} e^{i\pi \delta_{\alpha H} \delta_{\mu L}} + \sqrt{\frac{V_{\mu,\alpha}^*}{2\pi}} a_{\mu}^{\vphantom{\dag}}(k) c_{\alpha}^{\dag} e^{-i\pi \delta_{\alpha H} \delta_{\mu L}} \Biggr] dk.
\end{split}
\end{align}
The Hamiltonian $H$ conserves the total excitation number, i.e., it commutes with the total excitation number operator
\begin{equation}
N_E = \sum_{j=1,2} \sigma_j^+ \sigma_j^- + \sum_{\alpha=H,V} c_{\alpha}^{\dag} c_{\alpha}^{\vphantom{\dag}} + \sum_{\mu = L,R} \int a_{\mu}^{\dag}(k) a_{\mu}^{\vphantom{\dag}}(k) dk.
\end{equation}
This means that we can perform the transformation ${H \rightarrow H - \omega_0 N_E}$, as it is simply a constant energy shift that does not affect the dynamics of the system. After performing the transformation, we absorb the remaining factors of $\omega_0$ into the definitions of the emitter and cavity frequencies, i.e., ${\omega_{e,j} - \omega_0 \rightarrow \omega_{e,j}}$, and ${\omega_{c,\alpha} - \omega_0 \rightarrow \omega_{c,\alpha}}$. We then use the substitution ${k' = k + k_0}$ in the integrals containing $a_R(k)$, and ${k' = k - k_0}$ in the integrals containing $a_L(k)$. After these substitutions are used, we can relabel $k'$ with $k$, leading to:
\begin{align}
\begin{split}
H =& \sum_{j=1,2} \omega_{e,j}^{\vphantom{\dag}} \sigma_j^+ \sigma_j^- + \sum_{\alpha=H,V} \omega_{c,\alpha}^{\vphantom{\dag}} c_{\alpha}^{\dag} c_{\alpha}^{\vphantom{\dag}} - \int v_g k\; a_R^{\dag}(k - k_0) a_R^{\vphantom{\dag}}(k - k_0) dk + \int v_g k\; a_L^{\dag}(k + k_0) a_L^{\vphantom{\dag}}(k + k_0) dk\\[0.05in]
&+ \sum_{j=1,2} \sum_{\alpha = H,V} \Bigl( g_{\alpha,j}^{\vphantom{*}} \sigma_j^+ c_{\alpha}^{\vphantom{\dag}} + g_{\alpha,j}^* \sigma_j^- c_{\alpha}^\dag \Bigr) + \sum_{\alpha = H,V} \Biggl( \int \Biggl[ \sqrt{\frac{V_{R,\alpha}^{\vphantom{*}}}{2\pi}} a_R^{\dag}(k - k_0) c_{\alpha}^{\vphantom{\dag}} + \sqrt{\frac{V_{R,\alpha}^*}{2\pi}} a_R^{\vphantom{\dag}}(k - k_0) c_{\alpha}^{\dag} \Biggr]dk\\[0.05in]
&\hspace{2.75in} + \int \Biggl[ \sqrt{\frac{V_{L,\alpha}^{\vphantom{*}}}{2\pi}} a_L^{\dag}(k + k_0) c_{\alpha}^{\vphantom{\dag}} e^{i\pi \delta_{\alpha H}} + \sqrt{\frac{V_{L,\alpha}^*}{2\pi}} a_L^{\vphantom{\dag}}(k + k_0) c_{\alpha}^{\dag} e^{-i\pi \delta_{\alpha H}} \Biggr] dk \Biggr).
\end{split}
\end{align}
The transformation to frequency space is completed by defining the frequency variable ${\omega = -v_g k}$ for ${a_R(k-k_0)}$ integrals and ${\omega = v_g k}$ for integrals with ${a_L(k+k_0)}$. We also define the frequency-space waveguide operators ${a_R(\omega) = a_R(k-k_0)/\sqrt{v_g}}$ and ${a_L(\omega) = a_L(k+k_0)/\sqrt{v_g}}$, and absorb remaining factors of the group velocity into the cavity-waveguide coupling rates, i.e., ${V_{R,\alpha}/v_g \rightarrow V_{R,\alpha}}$ and ${V_{L,\alpha}/v_g \rightarrow V_{L,\alpha}}$. The resulting frequency-space Hamiltonian for the waveguide-coupled cavity is given in Eq.~(\ref{eq:H_freq}):
\begin{align}
\begin{split}
H =& \sum_{j=1,2} \omega_{e,j}^{\vphantom{\dag}} \sigma_j^+ \sigma_j^- + \sum_{\alpha=H,V} \omega_{c,\alpha}^{\vphantom{\dag}} c_{\alpha}^{\dag} c_{\alpha}^{\vphantom{\dag}} + \sum_{\mu = L,R} \int \omega a_{\mu}^{\dag}(\omega) a_{\mu}^{\vphantom{\dag}}(\omega) d\omega + \sum_{j=1,2} \sum_{\alpha = H,V} \Bigl( g_{\alpha,j}^{\vphantom{*}} \sigma_j^+ c_{\alpha}^{\vphantom{\dag}} + g_{\alpha,j}^* \sigma_j^- c_{\alpha}^\dag \Bigr)\\[0.05in]
&+ \sum_{\alpha = H,V} \sum_{\mu = L,R} \int \Biggl[ \sqrt{\frac{V_{\mu,\alpha}^{\vphantom{*}}}{2\pi}} a_{\mu}^{\dag}(\omega) c_{\alpha}^{\vphantom{\dag}} e^{i\pi \delta_{\alpha H} \delta_{\mu L}} + \sqrt{\frac{V_{\mu,\alpha}^*}{2\pi}} a_{\mu}^{\vphantom{\dag}}(\omega) c_{\alpha}^{\dag} e^{-i\pi \delta_{\alpha H} \delta_{\mu L}} \Biggr] d\omega.
\label{eq:H_freq}
\end{split}
\end{align}


\newpage
\subsection{Input-Output Relations}\label{subsec:transmission_input_output}

Using the Hamiltonian in Eq.~(\ref{eq:H_freq}), we can derive the input-output relations for our system. We begin by obtaining the Heisenberg equations for the time evolution of the waveguide operators $a_{\mu}(\omega,t)$:
\begin{equation}
\frac{d}{dt}a_{\mu}(\omega,t) = i\bigl[H, a_{\mu}(\omega,t)\bigr] = -i\omega a_{\mu}(\omega,t) - i \sum_{\alpha = H,V} e^{i\pi \delta_{\alpha H} \delta_{\mu L}} \sqrt{\frac{V_{\mu,\alpha}^{\vphantom{*}}}{2\pi}} c_{\alpha}^{\vphantom{\dag}}(t),
\end{equation}
where ${\mu \in \{L,R\}}$, and we have used the commutator ${[a_{\mu}^{\vphantom{\dag}}(\omega,t), a_{\mu}^{\dag}(\omega',t)] = \delta(\omega-\omega')}$ (all other equal-time commutators involving the waveguide operators are zero). Multiplying both sides by $e^{i\omega t}$ and rearranging gives
\begin{equation}
\frac{d}{dt}\Bigl[ a_{\mu}(\omega,t) e^{i\omega t} \Bigr] = -i \sum_{\alpha = H,V} e^{i\pi \delta_{\alpha H} \delta_{\mu L}} \sqrt{\frac{V_{\mu,\alpha}^{\vphantom{*}}}{2\pi}} c_{\alpha}^{\vphantom{\dag}}(t) e^{i\omega t}.
\label{eq:wg_Heisenberg}
\end{equation}
We now relabel $t$ with $t'$, and integrate from an `input time' $t_0$ to some time $t$:
\begin{equation}
a_{\mu}(\omega,t) e^{i\omega t} - a_{\mu}(\omega,t_0) e^{i\omega t_0} = -i \sum_{\alpha = H,V} e^{i\pi \delta_{\alpha H} \delta_{\mu L}} \sqrt{\frac{V_{\mu,\alpha}^{\vphantom{*}}}{2\pi}} \int_{t_0}^t c_{\alpha}^{\vphantom{\dag}}(t') e^{i\omega t'} dt'.
\end{equation}
Next, we multiply each term by $e^{-i\omega t}$, and then integrate over all $\omega$:
\begin{align}
\begin{split}
\int a_{\mu}(\omega,t) d\omega - \int a_{\mu}(\omega,t_0) e^{-i\omega(t-t_0)} d\omega &= -2 \pi i \sum_{\alpha = H,V} e^{i\pi \delta_{\alpha H} \delta_{\mu L}} \sqrt{\frac{V_{\mu,\alpha}^{\vphantom{*}}}{2\pi}} \int_{t_0}^t dt' c_{\alpha}^{\vphantom{\dag}}(t') \left[ \int \frac{d\omega}{2\pi} e^{i\omega(t'-t)} \right]\\[0.05in]
&= -2 \pi i \sum_{\alpha = H,V} e^{i\pi \delta_{\alpha H} \delta_{\mu L}} \sqrt{\frac{V_{\mu,\alpha}^{\vphantom{*}}}{2\pi}} \int_{t_0}^t dt' c_{\alpha}^{\vphantom{\dag}}(t') \delta(t'-t)\\[0.05in]
&= -i\pi \sum_{\alpha = H,V} e^{i\pi \delta_{\alpha H} \delta_{\mu L}} \sqrt{\frac{V_{\mu,\alpha}^{\vphantom{*}}}{2\pi}} c_{\alpha}^{\vphantom{\dag}}(t).
\end{split}
\end{align}
Here, we get a factor of ${1/2}$ from the integral with respect to $t'$ because ${\delta(t'-t)}$ is centered at one of the integration limits. Dividing through by $\sqrt{2\pi}$ leads to
\begin{equation}
\frac{1}{\sqrt{2\pi}}\int a_{\mu}(\omega,t) d\omega - a_{\mu,\text{in}}(t) = -\frac{i}{2} \sum_{\alpha = H,V} e^{i\pi \delta_{\alpha H} \delta_{\mu L}} \sqrt{V_{\mu,\alpha}^{\vphantom{*}}} c_{\alpha}^{\vphantom{\dag}}(t),
\label{eq:input}
\end{equation}
where
\begin{equation}
a_{\mu,\text{in}}(t) = \frac{1}{\sqrt{2\pi}}\int a_{\mu}(\omega,t_0) e^{-i\omega(t-t_0)} d\omega
\end{equation}
is the definition of an input operator in the input-output formalism~\cite{Gardiner1985}. We now return to Eq.~(\ref{eq:wg_Heisenberg}), relabel $t$ with $t'$, integrate from some time $t$ to an `output time' $t_1$, and repeat the remaining steps outlined above to get
\begin{equation}
a_{\mu,\text{out}}(t) - \frac{1}{\sqrt{2\pi}}\int a_{\mu}(\omega,t) d\omega = -\frac{i}{2} \sum_{\alpha = H,V} e^{i\pi \delta_{\alpha H} \delta_{\mu L}} \sqrt{V_{\mu,\alpha}^{\vphantom{*}}} c_{\alpha}^{\vphantom{\dag}}(t),
\label{eq:output}
\end{equation}
where
\begin{equation}
a_{\mu,\text{out}}(t) = \frac{1}{\sqrt{2\pi}}\int a_{\mu}(\omega,t_1) e^{-i\omega(t-t_1)} d\omega
\end{equation}
is the definition of an output operator in the input-output formalism~\cite{Gardiner1985}. From Eqs.~(\ref{eq:input}) and (\ref{eq:output}), it follows that the input and output operators for waveguide $\mu$ are given by:
\begin{subequations}
\begin{align}
a_{\mu,\text{in}}(t) &= \frac{1}{\sqrt{2\pi}} \int a_{\mu}(\omega,t) d\omega + \frac{i}{2} \sum_{\alpha = H,V} e^{i\pi \delta_{\alpha H} \delta_{\mu L}} \sqrt{V_{\mu,\alpha}^{\vphantom{*}}} c_{\alpha}^{\vphantom{\dag}}(t),\label{eq:input2}\\[0.05in]
a_{\mu,\text{out}}(t) &= \frac{1}{\sqrt{2\pi}} \int a_{\mu}(\omega,t) d\omega - \frac{i}{2} \sum_{\alpha = H,V} e^{i\pi \delta_{\alpha H} \delta_{\mu L}} \sqrt{V_{\mu,\alpha}^{\vphantom{*}}} c_{\alpha}^{\vphantom{\dag}}(t),\label{eq:output2}
\end{align}
\end{subequations}
which immediately lead to the input-output relations
\begin{equation}
a_{\mu,\text{out}}(t) = a_{\mu,\text{in}}(t) - i \sum_{\alpha = H,V} e^{i\pi \delta_{\alpha H} \delta_{\mu L}} \sqrt{V_{\mu,\alpha}^{\vphantom{*}}} c_{\alpha}^{\vphantom{\dag}}(t).
\label{eq:input_output_relations}
\end{equation}


\newpage
\subsection{Scattering Matrix}\label{subsec:transmission_scattering_matrix}

Using the input-output relations in Eq.~(\ref{eq:input_output_relations}), we can derive the single-photon scattering matrix~\cite{Fan2010}. As given in Appendix B, the scattering matrix elements have the form:
\begin{equation}
S_{pk}^{\mu\nu} = \expval{a_{\mu,\text{out}}(p) a_{\nu,\text{in}}^{\dag}(k)}{0},
\end{equation}
where $k$ is the frequency of the input photon in waveguide $\nu$, $p$ is the frequency of the output photon in waveguide $\mu$ (${\mu, \nu \in \{L,R\}}$), and ${\ket{0} = \ket{g} \otimes \ket{0_H} \otimes \ket{0_V} \otimes \ket{0_R} \otimes \ket{0_L}}$ is the vacuum state of the system. Using the Fourier transform of Eq.~(\ref{eq:input_output_relations}), we have:
\begin{align}
\begin{split}
S_{pk}^{\mu\nu} &= \expval{\Biggl[a_{\mu,\text{in}}(p) - i \sum_{\alpha = H,V} e^{i\pi \delta_{\alpha H} \delta_{\mu L}} \sqrt{V_{\mu,\alpha}^{\vphantom{*}}} c_{\alpha}^{\vphantom{\dag}}(p) \Biggr] a_{\nu,\text{in}}^{\dag}(k)}{0}\\[0.05in]
&= \expval{a_{\mu,\text{in}}(p) a_{\nu,\text{in}}^{\dag}(k)}{0} - i \sum_{\alpha = H,V} e^{i\pi \delta_{\alpha H} \delta_{\mu L}} \sqrt{V_{\mu,\alpha}^{\vphantom{*}}} \expval{c_{\alpha}^{\vphantom{\dag}}(p) a_{\nu,\text{in}}^{\dag}(k)}{0}\\[0.05in]
&= \delta_{\mu \nu} \delta(p-k) - i \sum_{\alpha = H,V} e^{i\pi \delta_{\alpha H} \delta_{\mu L}} \sqrt{V_{\mu,\alpha}^{\vphantom{*}}} \matrixel{0}{c_{\alpha}^{\vphantom{\dag}}(p)}{k_{\nu}^+},
\label{eq:S_matrix}
\end{split}
\end{align}
where we have used the commutator ${[a_{\mu,\text{in}}(p), a_{\nu,\text{in}}^{\dag}(k)] = \delta_{\mu \nu} \delta(p-k)}$, as well as the definition of the single-photon scattering eigenstate $\smash{\ket{k_{\nu}^+} = a_{\nu,\text{in}}^{\dag}(k)\ket{0}}$. We note that, in order to relate the input and output operators in the time domain (as defined in the input-output formalism) to the frequency-domain input and output operators in the scattering matrix through a Fourier transform, we formally need to take ${t_0 \rightarrow -\infty}$ as the input time and ${t_1 \rightarrow \infty}$ as the output time~\cite{Fan2010}. From Eq.~(\ref{eq:S_matrix}), we see that we need to calculate the matrix element $\matrixel{0}{c_{\alpha}^{\vphantom{\dag}}(p)}{k_{\nu}^+}$ for ${\alpha \in \{H,V\}}$ in order to obtain the scattering matrix. For this, we need to solve the Heisenberg equations for the cavity operators $c_{\alpha}^{\vphantom{\dag}}(t)$. If we include Lindblad terms corresponding to photon loss from cavity mode $\alpha$ at rate $\kappa_{\alpha}$ in the Heisenberg equations, we get:
\begin{align}
\begin{split}
\frac{d}{dt}c_{\alpha}^{\vphantom{\dag}}(t) =&\; i\bigl[H, c_{\alpha}^{\vphantom{\dag}}(t)\bigr] + L_{c,\alpha}^{\dag}c_{\alpha}^{\vphantom{\dag}}(t)L_{c,\alpha}^{\vphantom{\dag}} - \frac{1}{2} \Bigl\{L_{c,\alpha}^{\dag}L_{c,\alpha}^{\vphantom{\dag}},c_{\alpha}^{\vphantom{\dag}}(t)\Bigr\}\\[0.05in]
=& -i \left( \omega_{c,\alpha} - \frac{i\kappa_{\alpha}}{2} \right) c_{\alpha}(t) - i \sum_{j=1,2} g_{\alpha,j}^* \sigma_j^-(t) - i \sum_{\mu=L,R} e^{-i\pi \delta_{\alpha H} \delta_{\mu L}} \sqrt{\frac{V_{\mu,\alpha}^*}{2\pi}} \int a_{\mu}(\omega,t) d\omega\\[0.05in]
=& -i \left( \omega_{c,\alpha} - \frac{i\kappa_{\alpha}}{2} \right) c_{\alpha}(t) - i \sum_{j=1,2} g_{\alpha,j}^* \sigma_j^-(t) - i \sum_{\mu=L,R} e^{-i\pi \delta_{\alpha H} \delta_{\mu L}} \sqrt{V_{\mu,\alpha}^*} a_{\mu,\text{in}}(t)\\[0.05in]
&- \frac{1}{2} \sum_{\mu=L,R} \sum_{\beta=H,V} e^{i\pi \delta_{\mu L}(\delta_{\beta H} - \delta_{\alpha H})} \sqrt{V_{\mu,\beta}^{\vphantom{*}} V_{\mu,\alpha}^*} c_{\beta}^{\vphantom{\dag}}(t),
\end{split}
\end{align}
where ${L_{c,\alpha} = \sqrt{\kappa_{\alpha}} c_{\alpha}}$ is the Lindblad operator for loss from mode $\alpha$, and we used the commutator ${[c_{\alpha}^{\vphantom{\dag}}(t), c_{\alpha}^{\dag}(t)] = 1}$ (all other equal-time commutators involving the cavity operators are zero). We also used Eq.~(\ref{eq:input2}) to eliminate the integral in the second line. Note that adding the Lindblad terms is equivalent to replacing the cavity mode frequencies $\omega_{c,\alpha}$ with ${\omega_{c,\alpha} - i\kappa_{\alpha}/2}$, effectively making the frequencies complex. Next, we perform a Fourier transform from time $t$ to frequency $p$, allowing us to eliminate the time derivative:
\begin{align}
\begin{split}
-ip c_{\alpha}^{\vphantom{\dag}}(p) =& -i \left( \omega_{c,\alpha} - \frac{i\kappa_{\alpha}}{2} \right) c_{\alpha}(p) - i \sum_{j=1,2} g_{\alpha,j}^* \sigma_j^-(p) - i \sum_{\mu=L,R} e^{-i\pi \delta_{\alpha H} \delta_{\mu L}} \sqrt{V_{\mu,\alpha}^*} a_{\mu,\text{in}}(p)\\[0.05in]
&- \frac{1}{2} \sum_{\mu=L,R} \sum_{\beta=H,V} e^{i\pi \delta_{\mu L}(\delta_{\beta H} - \delta_{\alpha H})} \sqrt{V_{\mu,\beta}^{\vphantom{*}} V_{\mu,\alpha}^*} c_{\beta}^{\vphantom{\dag}}(p).
\end{split}
\end{align}
If we pre-multiply by $\bra{0}$ and post-multiply by $\ket{k_{\nu}^+}$, and use ${\matrixel{0}{a_{\mu,\text{in}}(p)}{k_{\nu}^+} = \expval{a_{\mu,\text{in}}(p) a_{\nu,\text{in}}^{\dag}(k)}{0} = \delta_{\mu \nu} \delta(p-k)}$, we obtain the following:
\begin{align}
\begin{split}
-i \left( \Delta_{c,\alpha} + \frac{i\kappa_{\alpha}}{2} \right) \matrixel{0}{c_{\alpha}^{\vphantom{\dag}}(p)}{k_{\nu}^+} =& - i \sum_{j=1,2} g_{\alpha,j}^* \matrixel{0}{\sigma_j^-(p)}{k_{\nu}^+} - i e^{-i\pi \delta_{\alpha H} \delta_{\nu L}} \sqrt{V_{\nu,\alpha}^*} \delta(p-k)\\[0.05in]
&- \frac{1}{2} \sum_{\mu=L,R} \sum_{\beta=H,V} e^{i\pi \delta_{\mu L}(\delta_{\beta H} - \delta_{\alpha H})} \sqrt{V_{\mu,\beta}^{\vphantom{*}} V_{\mu,\alpha}^*} \matrixel{0}{c_{\beta}^{\vphantom{\dag}}(p)}{k_{\nu}^+},
\label{eq:matrix_el_cavity}
\end{split}
\end{align}
where we have also defined the frequency detuning $\Delta_{c,\alpha} = p - \omega_{c,\alpha}$. In order to find the matrix elements $\matrixel{0}{c_H^{\vphantom{\dag}}(p)}{k_{\nu}^+}$ and $\matrixel{0}{c_V^{\vphantom{\dag}}(p)}{k_{\nu}^+}$ [and hence the scattering matrix elements in Eq.~(\ref{eq:S_matrix})], we therefore need to find $\matrixel{0}{\sigma_j^-(p)}{k_{\nu}^+}$ for ${j \in \{1,2\}}$. We do this using the Heisenberg equation for $\sigma_j^-(t)$, where we include Lindblad terms corresponding to photon loss from transition $j$ of the emitter at rate $\gamma_j$ (corresponding to emission into non-cavity modes):
\begin{align}
\begin{split}
\frac{d}{dt}\sigma_j^-(t) &= i\Bigl[H, \sigma_j^-(t)\Bigr] + L_{e,j}^{\dag}\sigma_j^-(t)L_{e,j}^{\vphantom{\dag}} - \frac{1}{2}\Bigl\{L_{e,j}^{\dag}L_{e,j}^{\vphantom{\dag}},\sigma_j^-(t)\Bigr\}\\[0.05in]
&= -i \left( \omega_{e,j} - \frac{i\gamma_j}{2} \right) \sigma_j^-(t) + i \sum_{k=1,2} \sum_{\alpha=H,V} g_{\alpha,k} e^{iHt} \ketbra{e_k}{e_j} e^{-iHt} c_{\alpha}^{\vphantom{\dag}}(t) - i \sum_{\alpha=H,V} g_{\alpha,j} e^{iHt} \ketbra{g}{g} e^{-iHt} c_{\alpha}^{\vphantom{\dag}}(t),
\label{eq:emitter_Heisenberg}
\end{split}
\end{align}
where ${L_{e,j} = \sqrt{\gamma_j} \sigma_j^-}$ is the Lindblad operator for loss from transition $j$, and we used the time-evolved operators $\sigma_j^-(t) = {e^{iHt} \sigma_j^-(0) e^{-iHt}} = {e^{iHt} \ketbra{g}{e_j} e^{-iHt}}$ and $\sigma_j^+(t) = {e^{iHt} \sigma_j^+(0) e^{-iHt}} = {e^{iHt} \ketbra{e_j}{g} e^{-iHt}}$ to evaluate the commutators and operator products. Note again that including the Lindblad terms is equivalent to replacing $\omega_{e,j}$ with ${\omega_{e,j} - i\gamma_j/2}$, effectively making the emitter transition frequencies complex. We now pre-multiply by $\bra{0}$ and post-multiply by $\ket{k_{\nu}^+}$ to obtain:
\begin{align}
\begin{split}
\frac{d}{dt}\matrixel{0}{\sigma_j^-(t)}{k_{\nu}^+} =& -i \left( \omega_{e,j} - \frac{i\gamma_j}{2} \right) \matrixel{0}{\sigma_j^-(t)}{k_{\nu}^+} + i \sum_{k=1,2} \sum_{\alpha=H,V} g_{\alpha,k} \bra{0}e^{iHt}\ket{e_k} \bra{e_j}e^{-iHt} c_{\alpha}^{\vphantom{\dag}}(t) |k_{\nu}^+ \rangle\\[0.05in]
&- i \sum_{\alpha=H,V} g_{\alpha,j} \bra{0}e^{iHt}\ket{g} \bra{g}e^{-iHt} c_{\alpha}^{\vphantom{\dag}}(t) |k_{\nu}^+ \rangle\\[0.05in]
=& -i \left( \omega_{e,j} - \frac{i\gamma_j}{2} \right) \matrixel{0}{\sigma_j^-(t)}{k_{\nu}^+} + i \sum_{k=1,2} \sum_{\alpha=H,V} g_{\alpha,k} \braket{0}{e_k} \bra{e_j}e^{-iHt} c_{\alpha}^{\vphantom{\dag}}(t) |k_{\nu}^+ \rangle\\[0.05in]
&- i \sum_{\alpha=H,V} g_{\alpha,j} \braket{0}{g} \bra{g}e^{-iHt} c_{\alpha}^{\vphantom{\dag}}(t) |k_{\nu}^+ \rangle\\[0.05in]
=& -i \left( \omega_{e,j} - \frac{i\gamma_j}{2} \right) \matrixel{0}{\sigma_j^-(t)}{k_{\nu}^+} - i \sum_{\alpha=H,V} g_{\alpha,j} \matrixel{0}{e^{-iHt} c_{\alpha}^{\vphantom{\dag}}(t)}{k_{\nu}^+}\\[0.05in]
=& -i \left( \omega_{e,j} - \frac{i\gamma_j}{2} \right) \matrixel{0}{\sigma_j^-(t)}{k_{\nu}^+} - i \sum_{\alpha=H,V} g_{\alpha,j} \matrixel{0}{c_{\alpha}^{\vphantom{\dag}}(t)}{k_{\nu}^+},
\end{split}
\end{align}
where we used ${e^{\pm iHt}\ket{0} = \ket{0}}$ (which follows from ${H \ket{0} = 0}$), as well as ${\braket{0}{e_k} = 0}$ and ${\braket{0}{g}\bra{g} = \bra{0}}$. As before, we Fourier transform from time to frequency to eliminate the time derivative and obtain an algebraic equation involving the matrix elements:
\begin{equation}
-ip \matrixel{0}{\sigma_j^-(p)}{k_{\nu}^+} = -i \left( \omega_{e,j} - \frac{i\gamma_j}{2} \right) \matrixel{0}{\sigma_j^-(p)}{k_{\nu}^+} - i \sum_{\alpha=H,V} g_{\alpha,j} \matrixel{0}{c_{\alpha}^{\vphantom{\dag}}(p)}{k_{\nu}^+}.
\end{equation}
Solving for $\matrixel{0}{\sigma_j^-(p)}{k_{\nu}^+}$ gives
\begin{equation}
\matrixel{0}{\sigma_j^-(p)}{k_{\nu}^+} = \frac{1}{\Delta_{e,j} + \frac{i\gamma_j}{2}} \sum_{\alpha=H,V} g_{\alpha,j} \matrixel{0}{c_{\alpha}^{\vphantom{\dag}}(p)}{k_{\nu}^+},
\label{eq:matrix_el_emitter}
\end{equation}
where we have defined the detuning $\Delta_{e,j} = p - \omega_{e,j}$. Substituting the result in Eq.~(\ref{eq:matrix_el_emitter}) into Eq.~(\ref{eq:matrix_el_cavity}) gives:
\begin{align}
\begin{split}
-i \left( \Delta_{c,\alpha} + \frac{i\kappa_{\alpha}}{2} \right) \matrixel{0}{c_{\alpha}^{\vphantom{\dag}}(p)}{k_{\nu}^+} =& - \sum_{\beta=H,V} \Biggl[ \hspace{0.02in} \sum_{j=1,2} \frac{i g_{\beta,j}^{\vphantom{*}} g_{\alpha,j}^*}{\Delta_{e,j} + \frac{i\gamma_j}{2}} + \frac{1}{2} \sum_{\mu=L,R} e^{i\pi \delta_{\mu L}(\delta_{\beta H} - \delta_{\alpha H})} \sqrt{V_{\mu,\beta}^{\vphantom{\dag}} V_{\mu,\alpha}^*} \Biggr] \matrixel{0}{c_{\beta}^{\vphantom{\dag}}(p)}{k_{\nu}^+}\\[0.05in]
&- i e^{-i\pi \delta_{\alpha H} \delta_{\nu L}} \sqrt{V_{\nu,\alpha}^*} \delta(p-k).
\label{eq:cavity_equations}
\end{split}
\end{align}
Since $\alpha$ is a free index, Eq.~(\ref{eq:cavity_equations}) consists of two simultaneous equations (for ${\alpha = H}$ and ${\alpha = V}$), which we can solve for the matrix elements $\matrixel{0}{c_H^{\vphantom{\dag}}(p)}{k_{\nu}^+}$ and $\matrixel{0}{c_V^{\vphantom{\dag}}(p)}{k_{\nu}^+}$. The results are given in Eqs.~(\ref{eq:CH}) and (\ref{eq:CV}) below:
\begin{subequations}
\begin{equation}
\matrixel{0}{c_H^{\vphantom{\dag}}(p)}{k_{\nu}^+} = \frac{e^{-i\pi \delta_{\nu L}} A_V \sqrt{V_{\nu,H}^*} - B_{V,H} \sqrt{V_{\nu,V}^*}}{A_H A_V - B_{H,V} B_{V,H}} \delta(p-k),
\label{eq:CH}
\end{equation}
\begin{equation}
\matrixel{0}{c_V^{\vphantom{\dag}}(p)}{k_{\nu}^+} = \frac{A_H \sqrt{V_{\nu,V}^*} - e^{-i\pi \delta_{\nu L}} B_{H,V} \sqrt{V_{\nu,H}^*}}{A_H A_V - B_{H,V} B_{V,H}} \delta(p-k),
\label{eq:CV}
\end{equation}
\end{subequations}
where we have defined
\begin{subequations}
\begin{equation}
A_{\alpha} = \Delta_{c,\alpha} + \frac{i\kappa_{\alpha}}{2} - \sum_{j=1,2} \frac{\left| g_{\alpha,j} \right|^2}{\Delta_{e,j} + \frac{i\gamma_j}{2}} + \frac{i}{2} \sum_{\mu=L,R} \left| V_{\mu,\alpha} \right|,
\end{equation}
\begin{equation}
B_{\alpha,\beta} = \frac{i}{2} \sum_{\mu=L,R} e^{i\pi \delta_{\mu L}(\delta_{\alpha H} - \delta_{\beta H})} \sqrt{V_{\mu,\alpha}^{\vphantom{*}} V_{\mu,\beta}^*} - \sum_{j=1,2} \frac{g_{\alpha,j}^{\vphantom{*}} g_{\beta,j}^*}{\Delta_{e,j} + \frac{i\gamma_j}{2}}.
\end{equation}
\end{subequations}
The scattering matrix elements are then found by substituting these results into Eq.~(\ref{eq:S_matrix}):
\begin{align}
\begin{split}
S_{pk}^{\mu \nu} =&\; \delta_{\mu \nu} \delta(p-k) - i e^{i\pi \delta_{\mu L}} \sqrt{V_{\mu,H}^{\vphantom{*}}} \matrixel{0}{c_{H}^{\vphantom{\dag}}(p)}{k_{\nu}^+} - i \sqrt{V_{\mu,V}^{\vphantom{*}}} \matrixel{0}{c_{V}^{\vphantom{\dag}}(p)}{k_{\nu}^+}\\[0.05in]
=& \left( \delta_{\mu \nu} + \frac{i \left[ e^{i\pi \delta_{\mu L}} B_{V,H} \sqrt{V_{\mu,H}^{\vphantom{*}} V_{\nu,V}^*} + e^{-i\pi \delta_{\nu L}} B_{H,V} \sqrt{V_{\mu,V}^{\vphantom{*}} V_{\nu,H}^*} - e^{i\pi (\delta_{\mu L} - \delta_{\nu L})} A_V \sqrt{V_{\mu,H}^{\vphantom{*}} V_{\nu,H}^*} - A_H \sqrt{V_{\mu,V}^{\vphantom{*}} V_{\nu,V}^*} \hspace{0.02in} \right]}{A_H A_V - B_{H,V} B_{V,H}} \right)\\
&\times \delta(p-k).
\end{split}
\end{align}

For transmission from the left waveguide to the right waveguide, we set ${\nu = L}$ and ${\mu = R}$ (since $\nu$ labels the waveguide of the input photon and $\mu$ labels the waveguide of the output photon). This gives us
\begin{equation}
S_{pk}^{RL} = t_{RL} \delta(p-k),
\end{equation}
where
\begin{equation}
t_{RL} = \frac{i \left( B_{V,H} \sqrt{V_{R,H}^{\vphantom{*}} V_{L,V}^*} - B_{H,V} \sqrt{V_{R,V}^{\vphantom{*}} V_{L,H}^*} + A_V \sqrt{V_{R,H}^{\vphantom{*}} V_{L,H}^*} - A_H \sqrt{V_{R,V}^{\vphantom{*}} V_{L,V}^*} \right)}{A_H A_V - B_{H,V} B_{V,H}}
\label{eq:t_RL}
\end{equation}
is the transmission coefficient for left-to-right (${L \rightarrow R}$) propagation. From this result, we can calculate the phase acquired by a transmitted photon, as this is given by the argument of the complex amplitude $t_{RL}$:
\begin{equation}
\phi_{RL} = \text{arg}\bigl(t_{RL}\bigr) = \text{atan2}\bigl(\Im\{t_{RL}\}, \Re\{t_{RL}\}\bigr),
\label{eq:phase_RL}
\end{equation}
where $\Re$ and $\Im$ denote the real and imaginary parts, respectively, and $\text{atan2}$ is the two-argument arctangent function.

By setting ${\nu = R}$ and ${\mu = L}$, we can also calculate the scattering matrix element for transmission from the right waveguide to the left waveguide:
\begin{equation}
S_{pk}^{LR} = t_{LR} \delta(p-k),
\end{equation}
where
\begin{equation}
t_{LR} = \frac{i \left( -B_{V,H} \sqrt{V_{L,H}^{\vphantom{*}} V_{R,V}^*} + B_{H,V} \sqrt{V_{L,V}^{\vphantom{*}} V_{R,H}^*} + A_V \sqrt{V_{L,H}^{\vphantom{*}} V_{R,H}^*} - A_H \sqrt{V_{L,V}^{\vphantom{*}} V_{R,V}^*} \right)}{A_H A_V - B_{H,V} B_{V,H}}
\label{eq:t_LR}
\end{equation}
is the transmission coefficient for right-to-left (${R \rightarrow L}$) propagation, from which we can calculate the corresponding transmission phase:
\begin{equation}
\phi_{LR} = \text{arg}\bigl(t_{LR}\bigr) = \text{atan2}\bigl(\Im\{t_{LR}\}, \Re\{t_{LR}\}\bigr).
\label{eq:phase_LR}
\end{equation}
The fits to the transmission data in the main text were obtained by calculating the transmission $|t_{RL}|^2$ with the experimentally measured system parameters (equivalently, we could calculate $|t_{LR}|^2$, as the cavity mode spectra from the main text are the same for ${L \rightarrow R}$ and ${R \rightarrow L}$ transmission). It is worth noting that, without the phases $e^{\pm i\pi \delta_{\alpha H} \delta_{\mu L}}$ in the Hamiltonian, the transmission results $t_{RL}$ and $t_{LR}$ would be invariant under the label exchange ${H \leftrightarrow V}$. This symmetry is broken by the phases, where one of the cavity modes couples anti-symmetrically into the waveguides, which is the origin of the directional emission from the cavity.


\newpage
\subsection{Comparison with FDTD Simulations}\label{subsec:transmission_simulations}

\begin{figure}[h!]
    \centering
    \includegraphics[width=0.5\textwidth]{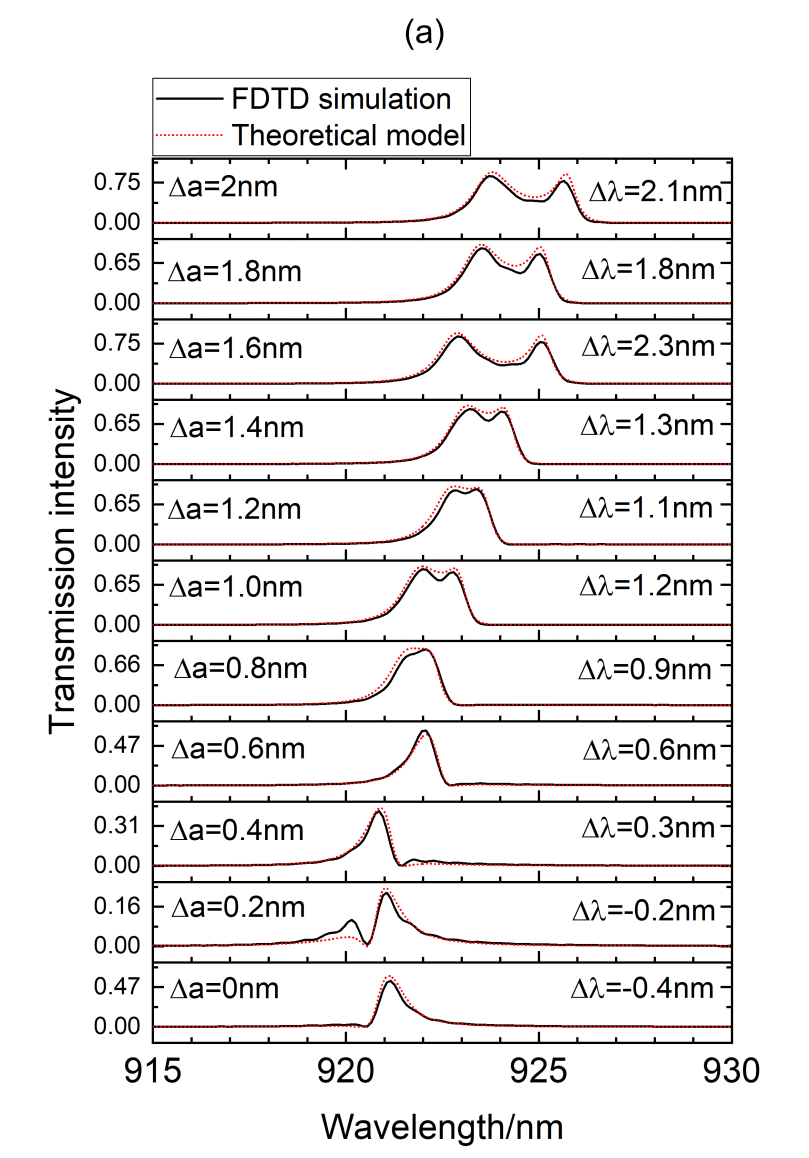}
    \caption{(a) Comparison between the FDTD simulated transmission (solid, black curves) and theoretical transmission (dashed, red curves) for the waveguide-coupled cavity. We consider different values of the photonic crystal lattice period difference ${\Delta a = a_y - a_x}$, giving rise to different mode detunings $\Delta \lambda$. The cavity parameters obtained from simulations were used to calculate the transmission using the scattering matrix model.}
    \label{fig:scatter}
\end{figure}


\newpage
\subsection{Resonant Transmission and Phase Shift Results}\label{subsec:transmission_resonant}

\begin{figure}[b!]
    \centering
    \includegraphics[width=0.8\textwidth]{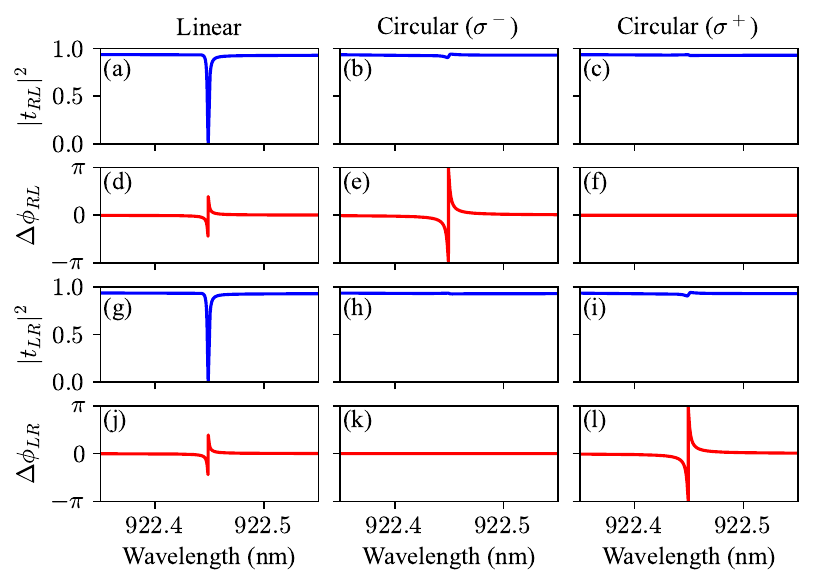}
    \caption{Transmission and phase shift results for a two-level emitter with transition wavelength ${\lambda_{e,1} = 922.45}$~nm in a waveguide-coupled cavity with mode wavelengths ${\lambda_{c,H} = 922.9}$~nm, ${\lambda_{c,V} = 921.9}$~nm, and $Q$ factors ${Q_H = 1600}$, ${Q_V = 700}$. We consider a linearly polarized emitter in the first column, a left-handed ($\sigma^-$) circularly polarized emitter in the second column, and a right-handed ($\sigma^+$) circularly polarized emitter in the third column. (a), (b), (c) show the transmission $|t_{RL}|^2$ for left-to-right propagation, and (d), (e), (f) show the corresponding phase shift ${\Delta \phi_{RL} = \phi_{RL} - \phi_{RL}|_{g=0}}$. (g), (h), (i) show the transmission $|t_{LR}|^2$ for right-to-left propagation, and (j), (k), (l) show the corresponding phase shift ${\Delta \phi_{LR} = \phi_{LR} - \phi_{LR}|_{g=0}}$.}
    \label{fig:transmission}
\end{figure}

In the main text, we use the scattering matrix model to fit experimental data corresponding to non-resonant transmission measurements. However, we can also use the model to predict how the transmission spectrum of our waveguide-coupled cavity device is influenced by resonant scattering from the emitter in the cavity. In Fig.~\ref{fig:transmission}, we show resonant transmission and phase shift results for three different emitter polarizations --- linear (first column), left-handed circular ($\sigma^-$, second column), and right-handed circular ($\sigma^+$, third column). The parameters we use for these results correspond to a simulated device~\cite{ChiralCav_2022}. In particular, the cavity mode wavelengths are ${\lambda_{c,H} = 922.9}$~nm and ${\lambda_{c,V} = 921.9}$~nm, and the quality ($Q$) factors of the modes are ${Q_H = 1600}$ and ${Q_V = 700}$, which we use to calculate the cavity-waveguide coupling rates ${V_{R,\alpha} = V_{L,\alpha} = \omega_{c,\alpha}/2Q_{\alpha}}$ for ${\alpha \in \{H,V\}}$ (${\omega_{c,\alpha} = 2\pi c/\lambda_{c,\alpha}}$, where $c$ is the vacuum speed of light). The cavity loss rates are ${\kappa_{\alpha} = \omega_{c,\alpha}/Q_{u,\alpha}}$, where ${Q_{u,\alpha} = 3 \times 10^4}$ are the mode $Q$ factors for an uncoupled cavity. For simplicity we consider a two-level emitter in the cavity, with a transition wavelength ${\lambda_{e,1} = 922.45}$~nm, where the directional contrast was predicted to be at a maximum for a circularly polarized emitter~\cite{ChiralCav_2022} (we neglect the other transition of the three-level emitter in the model by setting ${g_{H,2} = g_{V,2} = 0}$ for coupling between the cavity and transition $2$). We set ${\gamma_1 = \gamma_2 = 0}$ for the emitter loss rates to demonstrate the optimal performance of the device, in the absence of imperfect coupling efficiency. For linear polarization (first column in Fig.~\ref{fig:transmission}), we set ${g_{H,1}/2\pi = 10}$~GHz and ${g_{V,1} = 0}$, such that the emitter only couples to the $H$ mode. For $\sigma^-$ circular polarization (second column in Fig.~\ref{fig:transmission}), we use ${g_{H,1}/2\pi = 10}$~GHz and ${g_{V,1}/2\pi = -10i}$~GHz, and for $\sigma^+$ circular polarization (third column in Fig.~\ref{fig:transmission}), ${g_{H,1}/2\pi = 10}$~GHz and ${g_{V,1}/2\pi = 10i}$~GHz (i.e., the emitter couples to the $H$ and $V$ modes with equal strength and a $\pi/2$ phase difference).

In Figs.~\ref{fig:transmission}(a), (b), and (c), we show the transmission $|t_{RL}|^2$ for left-to-right propagation [calculated using Eq.~(\ref{eq:t_RL})]. The corresponding transmission phase shifts ${\Delta \phi_{RL} = \phi_{RL} - \phi_{RL}|_{g=0}}$ are shown in Figs.~\ref{fig:transmission}(d), (e), and (f), which we calculate by taking the difference between the phase $\phi_{RL}$ [Eq.~(\ref{eq:phase_RL})] evaluated with the parameters given above and the same phase but with all the emitter-cavity coupling rates set to zero (in this way, we observe the phase shift due to the interaction with the emitter in the cavity). Similarly, Figs.~\ref{fig:transmission}(g), (h), and (i) show the transmission $|t_{LR}|^2$ for right-to-left propagation [Eq.~(\ref{eq:t_LR})], and the corresponding phase shifts ${\Delta \phi_{LR} = \phi_{LR} - \phi_{LR}|_{g=0}}$ are shown in Figs.~\ref{fig:transmission}(j), (k), and (l) [calculated using Eq.~(\ref{eq:phase_LR})]. We observe that, for a linearly polarized emitter (first column in Fig.~\ref{fig:transmission}), the transmission and phase shift are reciprocal, i.e., the same for left-to-right and right-to-left propagation. For both propagation directions, the transmission falls to zero on resonance with the emitter, and a moderate phase shift is acquired by the scattered photon. In contrast, we see that for a circularly polarized emitter, the phase shift is nonreciprocal, i.e., different for left-to-right and right-to-left propagation. For $\sigma^-$ circular polarization (second column in Fig.~\ref{fig:transmission}), a $\pi$ phase shift occurs on resonance for left-to-right propagation, but there is no phase shift for right-to-left propagation. The situation is reversed for $\sigma^+$ circular polarization (third column in Fig.~\ref{fig:transmission}), where a $\pi$ phase shift occurs for right-to-left propagation, but there is no phase shift for left-to-right propagation. Furthermore, for both $\sigma^+$ and $\sigma^-$ polarization, the transmission on resonance is almost perfect; there is minimal reflection as coupling occurs almost exclusively in a single direction. These results demonstrate that our cavity device can be used as a nonreciprocal phase shifter when a circularly polarized emitter is embedded in the cavity, where the nonreciprocity arises from the directional light–matter interaction within the cavity. Ideally, for one input direction we achieve a unidirectional interaction and a $\pi$ phase shift as a result, and for the other input direction there is no interaction and hence no phase shift, as coupling to one of the propagation directions is suppressed.


\newpage
\section{Quantum Trajectory Model for Emission Directionality}

In this section, we present the quantum trajectory model that we use to predict the directional contrast of our system. As mentioned in the main text, this model is based on a space-discretized waveguide approach~\cite{Regidor2021}. First, in Section~\ref{subsec:trajectories_Hamiltonian} we discretize our Hamiltonian from the main text [see Eq.~(\ref{eq:H_k})]. We then outline how this discretization allows us to simulate the time evolution of our system using quantum trajectory theory in Section~\ref{subsec:trajectories_algorithm}. Some example trajectory results are shown in Section~\ref{subsec:trajectories_results}, corresponding to Device~1 and Device~2 from the main text.


\subsection{Hamiltonian Discretization}\label{subsec:trajectories_Hamiltonian}

\begin{figure}[h]
    \includegraphics[width = 0.75\textwidth]{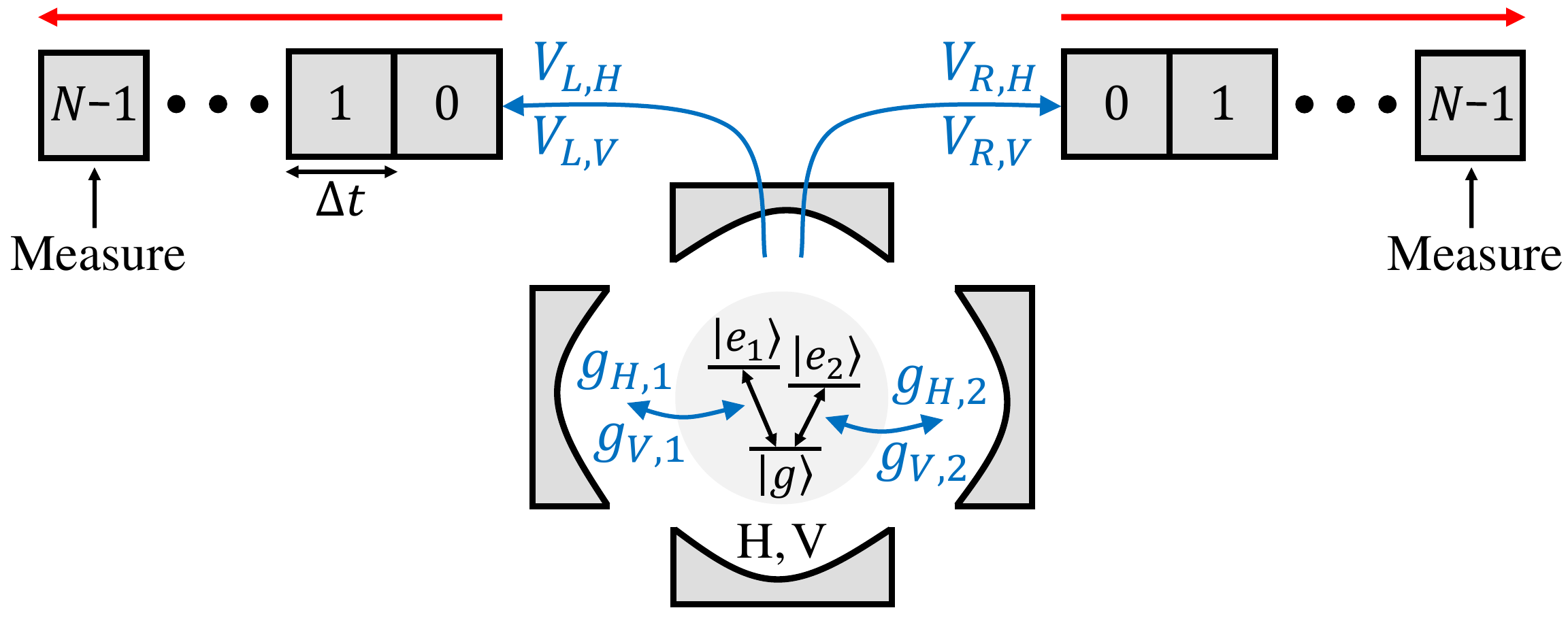}
    \caption{The waveguide-coupled cavity system treated within the space-discretized quantum trajectory model. Each waveguide is divided into $N$ spatial bins of width $\Delta t$ in time. The boxes are labeled with the index ${n \in \{0,1,\dotsc,N-1\}}$. At each time step in the quantum trajectory algorithm, the cavity interacts with box $0$ in each waveguide, photon number measurements are simulated on the final waveguide boxes, and the boxes are moved along by one (in the direction indicated by the red arrows).}
    \label{fig:discretized_waveguides}
\end{figure}

As mentioned in the main text, each of the waveguides is discretized into a series of $N$ spatial bins/boxes, see Fig.~\ref{fig:discretized_waveguides}. This discretization enables the time evolution of the system to be solved numerically using quantum trajectories. In order to discretize the $k$-space Hamiltonian given in Eq.~(\ref{eq:H_k}), we replace the integrals over $k$ with summations over a finite number $N$ of $k$-space modes, and we replace the continuum waveguide mode operators $\smash{a_{\mu}^{\vphantom{\dag}}(k)}$, $\smash{a_{\mu}^{\dag}(k)}$ with discrete mode operators $\smash{a_{\mu,k}^{\vphantom{\dag}}}$, $\smash{a_{\mu,k}^{\dag}}$~\cite{Regidor2021}:
\begin{align}
\begin{split}
H \rightarrow& \sum_{j=1,2} \omega_{e,j}^{\vphantom{\dag}} \sigma_j^+ \sigma_j^- + \sum_{\alpha=H,V} \omega_{c,\alpha}^{\vphantom{\dag}} c_{\alpha}^{\dag} c_{\alpha}^{\vphantom{\dag}} + \sum_{\mu = L,R} \sum_{k=0}^{N-1} \omega_k^{\vphantom{\dag}} a_{\mu,k}^{\dag} a_{\mu,k}^{\vphantom{\dag}} + \sum_{j=1,2} \sum_{\alpha = H,V} \Bigl( g_{\alpha,j}^{\vphantom{*}} \sigma_j^+ c_{\alpha}^{\vphantom{\dag}} + g_{\alpha,j}^* \sigma_j^- c_{\alpha}^\dag \Bigr)\\[0.05in]
&+ \sqrt{\frac{2\pi}{L_0}} \sum_{\alpha = H,V} \sum_{\mu = L,R} \sum_{k=0}^{N-1} \Biggl( \sqrt{\frac{V_{\mu,\alpha}^{\vphantom{*}}}{2\pi}} a_{\mu,k}^{\dag} c_{\alpha}^{\vphantom{\dag}} e^{i\pi \delta_{\alpha H} \delta_{\mu L}} + \sqrt{\frac{V_{\mu,\alpha}^*}{2\pi}} a_{\mu,k}^{\vphantom{\dag}} c_{\alpha}^{\dag} e^{-i\pi \delta_{\alpha H} \delta_{\mu L}} \Biggr),
\label{eq:H_discrete}
\end{split}
\end{align}
where ${L_0 = N \Delta t}$ is the length of the discretized waveguides. This allows us to perform discrete Fourier transforms that replace the $k$-space waveguide operators with position-space operators describing the spatial bins (see Section~\ref{subsec:trajectories_algorithm}):
\begin{equation}
A_{R,n} = \frac{1}{\sqrt{N}} \sum_{k=0}^{N-1} a_{R,k} e^{i \omega_k n \Delta t}, \quad \text{and} \quad A_{L,n} = \frac{1}{\sqrt{N}} \sum_{k=0}^{N-1} a_{L,k} e^{-i \omega_k n \Delta t},
\label{eq:FT}
\end{equation}
as given in Appendix C.

In order to perform numerical simulations of our system, we need to express the Hamiltonian (and any other observable of interest) as a matrix by choosing an appropriate basis. Since we look at single-photon emission from the cavity, the basis will only contain states with at most one photon in the system (higher photon number states are never occupied in this situation). At any given time, the photon can either exist as an excitation in the three-level emitter ($\ket{e_1}$ or $\ket{e_2}$), one of the two cavity modes, or one of the $2N$ spatial bins of the waveguides. Including also the ground state of the system, this amounts to a basis with ${2N+5}$ elements, as given in Eq.~(14) in Appendix C. As there are ${2N+5}$ basis elements, all observables are represented by ${(2N+5) \times (2N+5)}$ matrices, and the state of the system is represented by a vector of length ${2N+5}$.


\newpage
\subsection{Quantum Trajectory Algorithm}\label{subsec:trajectories_algorithm}

After discretizing the Hamiltonian, we can use it to solve for the time evolution of the system numerically. This allows us to calculate the probability that an emitted photon ends up in the left waveguide or the right waveguide, from which we can extract the directional contrast. To obtain the time evolution, we follow the quantum trajectory approach presented in ref.~\cite{Regidor2021}. First, we divide the Hamiltonian in Eq.~(\ref{eq:H_discrete}) into three parts:
\begin{equation}
H = H_S + H_W + H_I,
\end{equation}
where
\begin{equation}
H_S = \sum_{j=1,2} \omega_{e,j}^{\vphantom{\dag}} \sigma_j^+ \sigma_j^- + \sum_{\alpha = H,V} \omega_{c,\alpha}^{\vphantom{\dag}} c_{\alpha}^\dag c_{\alpha}^{\vphantom{\dag}} + \sum_{j=1,2} \sum_{\alpha = H,V} \Bigl( g_{\alpha,j}^{\vphantom{*}} \sigma_j^+ c_{\alpha}^{\vphantom{\dag}} + g_{\alpha,j}^* \sigma_j^- c_{\alpha}^\dag \Bigr)
\label{eq:H_S}
\end{equation}
is the Hamiltonian for the emitter-cavity system,
\begin{equation}
H_W = \sum_{\mu = L,R} \sum_{k=0}^{N-1} \omega_k^{\vphantom{\dag}} a_{\mu,k}^{\dag} a_{\mu,k}^{\vphantom{\dag}}
\label{eq:H_W}
\end{equation}
is the free waveguide Hamiltonian, and
\begin{equation}
H_I = \sqrt{\frac{2\pi}{L_0}} \sum_{\alpha = H,V} \sum_{\mu = L,R} \sum_{k=0}^{N-1} \Biggl( \sqrt{\frac{V_{\mu,\alpha}^{\vphantom{*}}}{2\pi}} a_{\mu,k}^{\dag} c_{\alpha}^{\vphantom{\dag}} e^{i\pi \delta_{\alpha H} \delta_{\mu L}} + \sqrt{\frac{V_{\mu,\alpha}^*}{2\pi}} a_{\mu,k}^{\vphantom{\dag}} c_{\alpha}^{\dag} e^{-i\pi \delta_{\alpha H} \delta_{\mu L}} \Biggr)
\label{eq:H_I}
\end{equation}
is the interaction between the cavity and the waveguides. As mentioned in Appendix C, we set the initial system state $\ket{\psi(0)}$ to the case where the emitter in the cavity is excited in some superposition of its excited states, $\ket{e_1}$ and $\ket{e_2}$:
\begin{equation}
\ket{\psi(0)} = \alpha \ket{e_1, 0_H, 0_V, 0_{R,N-1}, \dotsc, 0_{R,0}, 0_{L,N-1}, \dotsc, 0_{L,0}} + \beta \ket{e_2, 0_H, 0_V, 0_{R,N-1}, \dotsc, 0_{R,0}, 0_{L,N-1}, \dotsc, 0_{L,0}},
\label{eq:state}
\end{equation}
where ${|\alpha|^2 + |\beta|^2 = 1}$. Since $\alpha$ is the coefficient of basis element $2$ and $\beta$ is the coefficient of basis element $3$ [see Eq.~(14) in the Appendix], the initial state is represented by a vector where the second element is $\alpha$ and the third element is $\beta$, and all other elements are zero. The time evolution of the state $\ket{\psi(t)}$ is then obtained by following the algorithm below at each time step:\\
\\
\indent
(1) Calculate observables of interest at the beginning of the time step [for observable $O$, compute the expectation value ${\expval{O} = \expval{O}{\psi(t)}}$ for the current state $\ket{\psi(t)}$].\\
\\
\indent
(2) Evolve the state $\ket{\psi(t)}$ by one time step. If a loss occurs, apply the corresponding Lindblad jump operator $L_k$ and renormalize:
\begin{equation}
\ket{\psi(t+\Delta t)} = \frac{L_k^{\vphantom{\dag}} \ket{\psi(t)}}{\sqrt{\expval{L_k^{\dag} L_k^{\vphantom{\dag}}}{\psi(t)}}}.
\label{eq:loss}
\end{equation}
If a loss does not occur, evolve the state under the effective non-Hermitian Hamiltonian ${H_{\text{eff}} = H_S + H_I - i/2 \sum_k L_k^{\dag} L_k^{\vphantom{\dag}}}$ by directly applying the time evolution operator ${U_{\text{eff}}(\Delta t) = e^{-iH_{\text{eff}}\Delta t}}$ [and renormalize as $U_{\text{eff}}(\Delta t)$ is not unitary when losses are included in the model]:
\begin{equation}
\ket{\psi(t+\Delta t)} = \frac{U_{\text{eff}}^{\vphantom{\dag}}(\Delta t) \ket{\psi(t)}}{\sqrt{\expval{U_{\text{eff}}^{\dag}(\Delta t) U_{\text{eff}}^{\vphantom{\dag}}(\Delta t)}{\psi(t)}}}.
\label{eq:no_loss}
\end{equation}

(3) Simulate a photon number measurement on the final waveguide boxes, and project the state according to the measurement outcome.\\
\\
\indent
(4) Shift the waveguide boxes along by one. This corresponds to evolving the system under the free waveguide Hamiltonian $H_W$ in Eq.~(\ref{eq:H_W}) for a time $\Delta t$.\\
\\
\indent
(5) Renormalize the state before the next iteration:
\begin{equation}
\ket{\psi(t+\Delta t)} \rightarrow \frac{\ket{\psi(t+\Delta t)}}{\sqrt{\braket{\psi(t+\Delta t)}}}.
\label{eq:normalisation}
\end{equation}

In step~(1), we calculate the populations of the final waveguide boxes (labeled with index ${N-1}$, see Fig.~\ref{fig:discretized_waveguides}), as these give the probabilities for the photon being emitted into the respective waveguides. These populations are calculated using the position-space waveguide operators as (for ${\mu \in \{L,R\}}$):
\begin{equation}
\bigl\langle A_{\mu,N-1}^{\dag} A_{\mu,N-1}^{\vphantom{\dag}} \bigr\rangle = \expval{A_{\mu,N-1}^{\dag} A_{\mu,N-1}^{\vphantom{\dag}}}{\psi(t)}.
\label{eq:wg_pop}
\end{equation}
In addition to the waveguide box populations, we calculate the populations of the excited states $\ket{e_j}$ of the emitter (for ${j \in \{1,2\}}$),
\begin{equation}
\expval{\sigma_j^+ \sigma_j^-} = \expval{\sigma_j^+ \sigma_j^-}{\psi(t)} = \left| \braket{e_j}{\psi(t)} \right|^2,
\label{eq:emitter_pop}
\end{equation}
as well as the populations of the two cavity modes (${\alpha \in \{H,V\}}$):
\begin{equation}
\expval{c_{\alpha}^{\dag} c_{\alpha}^{\vphantom{\dag}}} = \expval{c_{\alpha}^{\dag} c_{\alpha}^{\vphantom{\dag}}}{\psi(t)}.
\label{eq:cavity_pop}
\end{equation}

In order to implement step~(2), we first need to determine if a loss occurs in the current time step. The losses we consider are the same as in the scattering matrix model, described by the Lindblad jump operators
\begin{equation}
L_{c,\alpha} = \sqrt{\kappa_{\alpha}} c_{\alpha} \quad \text{and} \quad L_{e,j} = \sqrt{\gamma_j} \sigma_j^-,
\label{eq:Lindblad}
\end{equation}
for ${\alpha \in \{H,V\}}$ (cavity losses) and ${j \in \{1,2\}}$ (emitter losses). We simulate whether a loss occurs by computing the total loss probability
\begin{equation}
P_{\text{loss}}(t) = \Delta t \sum_k \expval{L_k^{\dag} L_k^{\vphantom{\dag}}}{\psi(t)},
\end{equation}
and comparing it to a uniformly-distributed random number between $0$ and $1$, ${\epsilon \in (0,1)}$. If ${\epsilon \leq P_{\text{loss}}}$ then a loss occurs, and we evolve the state according to Eq.~(\ref{eq:loss}). We choose which of the Lindblad jump operators $L_k$ to apply by comparing the individual loss probabilities [given by ${P_k = \Delta t \expval{L_k^{\dag} L_k^{\vphantom{\dag}}}{\psi(t)}}$] against uniformly-distributed random numbers. On the other hand, if ${\epsilon > P_{\text{loss}}}$ then no loss occurs, and we evolve the state under the effective Hamiltonian ${H_{\text{eff}} = H_S + H_I - i/2 \sum_k L_k^{\dag} L_k^{\vphantom{\dag}}}$ [Eq.~(\ref{eq:no_loss})].

To be able to implement step~(2), we need to find matrix representations for the Lindblad operators $L_k$ and the time evolution operator ${U_{\text{eff}}(\Delta t) = e^{-iH_{\text{eff}}\Delta t}}$ in the chosen basis, such that they can act on the vector representation of $\ket{\psi(t)}$. For the Lindblad operators [Eq.~(\ref{eq:Lindblad})] and the emitter-cavity Hamiltonian $H_S$ [Eq.~(\ref{eq:H_S})] that goes into $U_{\text{eff}}(\Delta t)$, the matrix elements can immediately be found using the way in which the emitter operators $\sigma_j^\pm$ and the cavity operators $\smash{c_{\alpha}^{\vphantom{\dag}}}$, $\smash{c_{\alpha}^{\dag}}$ act on the basis states, as well as using the orthonormality of the basis elements. What then remains is to find the matrix for the cavity-waveguide interaction Hamiltonian $H_I$ [Eq.~(\ref{eq:H_I})], as this is also needed to find the matrix for $U_{\text{eff}}(\Delta t)$. Here, the $k$-space waveguide operators must first be replaced with the position-space operators for the spatial bins using the inverse of the Fourier transforms in Eq.~(\ref{eq:FT}), i.e.,
\begin{equation}
a_{R,k} = \frac{1}{\sqrt{N}} \sum_{n=0}^{N-1} A_{R,n} e^{-i \omega_k n \Delta t}, \quad \text{and} \quad a_{L,k} = \frac{1}{\sqrt{N}} \sum_{n=0}^{N-1} A_{L,n} e^{i \omega_k n \Delta t}.
\label{eq:inverse_FT}
\end{equation}
Substituting these into the $k$-space Hamiltonian in Eq.~(\ref{eq:H_I}) leads to the following:
\begin{align}
\begin{split}
H_I &=\; \sqrt{\frac{2\pi}{L_0}} \sum_{\alpha = H,V} \sum_{k=0}^{N-1} \Biggl[ \sqrt{\frac{V_{R,\alpha}^{\vphantom{*}}}{2\pi}} \biggl( \frac{1}{\sqrt{N}} \sum_{n=0}^{N-1} A_{R,n}^{\dag} e^{i \omega_k n \Delta t} \biggr) c_{\alpha}^{\vphantom{\dag}} + \sqrt{\frac{V_{R,\alpha}^*}{2\pi}} \biggl( \frac{1}{\sqrt{N}} \sum_{n=0}^{N-1} A_{R,n}^{\vphantom{\dag}} e^{-i \omega_k n \Delta t} \biggr) c_{\alpha}^{\dag} \\[0.05in]
&\hspace{1.1in}+ \sqrt{\frac{V_{L,\alpha}^{\vphantom{*}}}{2\pi}} \biggl( \frac{1}{\sqrt{N}} \sum_{n=0}^{N-1} A_{L,n}^{\dag} e^{-i \omega_k n \Delta t} \biggr) c_{\alpha}^{\vphantom{\dag}} e^{i\pi \delta_{\alpha H}} + \sqrt{\frac{V_{L,\alpha}^*}{2\pi}} \biggl( \frac{1}{\sqrt{N}} \sum_{n=0}^{N-1} A_{L,n}^{\vphantom{\dag}} e^{i \omega_k n \Delta t} \biggr) c_{\alpha}^{\dag} e^{-i\pi \delta_{\alpha H}} \Biggr]\\[0.05in]
&= \sqrt{\frac{2\pi}{N L_0}} \sum_{\alpha = H,V} \sum_{n=0}^{N-1} \Biggl[ \sqrt{\frac{V_{R,\alpha}^{\vphantom{*}}}{2\pi}} A_{R,n}^{\dag} c_{\alpha}^{\vphantom{\dag}} \biggl( \sum_{k=0}^{N-1}  e^{i \omega_k n \Delta t} \biggr) + \sqrt{\frac{V_{R,\alpha}^*}{2\pi}} A_{R,n}^{\vphantom{\dag}} c_{\alpha}^{\dag} \biggl( \sum_{k=0}^{N-1}  e^{-i \omega_k n \Delta t} \biggr)\\[0.05in]
&\hspace{1.25in}+ \sqrt{\frac{V_{L,\alpha}^{\vphantom{*}}}{2\pi}} A_{L,n}^{\dag} c_{\alpha}^{\vphantom{\dag}} e^{i\pi \delta_{\alpha H}} \biggl( \sum_{k=0}^{N-1} e^{-i \omega_k n \Delta t} \biggr) + \sqrt{\frac{V_{L,\alpha}^*}{2\pi}} A_{L,n}^{\vphantom{\dag}} c_{\alpha}^{\dag} e^{-i\pi \delta_{\alpha H}} \biggl( \sum_{k=0}^{N-1} e^{i \omega_k n \Delta t} \biggr) \Biggr].
\end{split}
\end{align}
Using the (discrete) linear dispersion approximation, ${\omega_k = 2\pi k/L_0}$, means that
\begin{equation}
\sum_{k=0}^{N-1} e^{\pm i \omega_k n \Delta t} = \sum_{k=0}^{N-1} e^{\pm 2\pi i k n \Delta t / L_0} = \sum_{k=0}^{N-1} e^{\pm 2\pi i k n / N} = N \left[ \frac{1}{N} \sum_{k=0}^{N-1} e^{\pm 2\pi i k (n-0) / N} \right] = N \delta_{n0},
\label{eq:exp_sum}
\end{equation}
where we have used ${L_0 = N \Delta t}$, as well as the identity
\begin{equation}
\delta_{nm} = \frac{1}{N} \sum_{k=0}^{N-1} e^{\pm 2\pi i k (n-m) / N}.
\end{equation}
Substituting the result in Eq.~(\ref{eq:exp_sum}) into $H_I$ allows the sum over $n$ to be eliminated, leading to:
\begin{align}
\begin{split}
H_I &=\; \sqrt{\frac{2\pi}{N L_0}} \sum_{\alpha = H,V} \sum_{n=0}^{N-1} N \delta_{n0} \Biggl( \sqrt{\frac{V_{R,\alpha}^{\vphantom{*}}}{2\pi}} A_{R,n}^{\dag} c_{\alpha}^{\vphantom{\dag}} + \sqrt{\frac{V_{R,\alpha}^*}{2\pi}} A_{R,n}^{\vphantom{\dag}} c_{\alpha}^{\dag} + \sqrt{\frac{V_{L,\alpha}^{\vphantom{*}}}{2\pi}} A_{L,n}^{\dag} c_{\alpha}^{\vphantom{\dag}} e^{i\pi \delta_{\alpha H}} + \sqrt{\frac{V_{L,\alpha}^*}{2\pi}} A_{L,n}^{\vphantom{\dag}} c_{\alpha}^{\dag} e^{-i\pi \delta_{\alpha H}} \Biggr)\\[0.05in]
&= \sqrt{\frac{N}{L_0}} \sum_{\alpha = H,V} \biggl( \sqrt{V_{R,\alpha}^{\vphantom{*}}} A_{R,0}^{\dag} c_{\alpha}^{\vphantom{\dag}} + \sqrt{V_{R,\alpha}^*} A_{R,0}^{\vphantom{\dag}} c_{\alpha}^{\dag} + \sqrt{V_{L,\alpha}^{\vphantom{*}}} A_{L,0}^{\dag} c_{\alpha}^{\vphantom{\dag}} e^{i\pi \delta_{\alpha H}} + \sqrt{V_{L,\alpha}^*} A_{L,0}^{\vphantom{\dag}} c_{\alpha}^{\dag} e^{-i\pi \delta_{\alpha H}} \biggr)\\[0.05in]
&= \sum_{\alpha = H,V} \sum_{\mu = L,R} \Biggl( \sqrt{\frac{V_{\mu,\alpha}^{\vphantom{*}}}{\Delta t}} A_{\mu,0}^{\dag} c_{\alpha}^{\vphantom{\dag}} e^{i\pi \delta_{\alpha H} \delta_{\mu L}} + \sqrt{\frac{V_{\mu,\alpha}^*}{\Delta t}} A_{\mu,0}^{\vphantom{\dag}} c_{\alpha}^{\dag} e^{-i\pi \delta_{\alpha H} \delta_{\mu L}} \Biggr).
\label{eq:H_I_final}
\end{split}
\end{align}
The matrix elements of $H_I$ can now be found using the way in which the cavity operators and the position-space waveguide operators act on the basis elements in Eq.~(14) in Appendix C (e.g., using that ${c_{\alpha}\ket{1_{\alpha}} = \ket{0_{\alpha}}}$, ${c_{\alpha}^{\dag}\ket{0_{\alpha}} = \ket{1_{\alpha}}}$, ${A_{\mu,n}\ket{1_{\mu,n}} = \ket{0_{\mu,n}}}$, and ${A_{\mu,n}^{\dag}\ket{0_{\mu,n}} = \ket{1_{\mu,n}}}$). Note that this Hamiltonian describes a local interaction between the cavity and the waveguide boxes labeled with ${n=0}$ (see Fig.~\ref{fig:discretized_waveguides}), which arises from the fact that the cavity-waveguide coupling is uniform in $k$-space (since it was assumed to be independent of $k$ through the Markov approximation~\cite{Gardiner1985}). 

In step~(3), a photon number measurement is simulated on the final waveguide boxes to determine whether a photon was detected at one of the waveguide ends in a given time step. First, the probabilities of measuring a photon in one of the final boxes must be calculated:
\begin{equation}
P_\mu = \braket{\psi_\mu},
\end{equation}
where $\ket{\psi_\mu}$ is the projected state where all amplitudes except that of the basis state containing $\ket{1_{\mu,N-1}}$ are set to zero (corresponding to a photon being present in the final box of waveguide $\mu$). For ${\mu=R}$ this is basis element $6$ and for ${\mu=L}$ this is basis element ${N+6}$. $P_\mu$ is then the probability that a photon is detected at the end of waveguide $\mu$ in the current time step. Next, the total detection probability ${P_{\text{tot}} = P_R + P_L}$ is compared against a uniformly-distributed random number ${\epsilon' \in (0,1)}$, in the same manner as losses are simulated in step~(2). If ${\epsilon' \leq P_{\text{tot}}}$, then a photon is detected at the end of one of the waveguides, and the current state is projected as
\begin{equation}
\ket{\psi(t+\Delta t)} \rightarrow \frac{\ket{\psi_\mu}}{\sqrt{\braket{\psi_\mu}}},
\end{equation}
where the waveguide in which the photon is detected (${\mu = R}$ or ${\mu = L}$) is determined by comparing the relative detection probabilities $P_R$ and $P_L$ against a second uniformly-distributed random number, ${\epsilon'' \in (0, P_{\text{tot}})}$ (if ${\epsilon'' \leq P_R}$ project onto $\ket{\psi_R}$, else project onto $\ket{\psi_L}$). On the other hand, if ${\epsilon' > P_{\text{tot}}}$, then no photon is detected and the state projection is:
\begin{equation}
\ket{\psi(t+\Delta t)} \rightarrow \frac{\ket{\psi_0}}{\sqrt{\braket{\psi_0}}},
\end{equation}
where $\ket{\psi_0}$ is the projected state with the amplitudes of the basis states containing $\ket{1_{R,N-1}}$ and $\ket{1_{L,N-1}}$ set to zero (elements $6$ and ${N+6}$).

In step~(4), the waveguide boxes are shifted along by one. This is implemented by moving the amplitudes in the vector ${\ket{\psi(t+\Delta t)}}$ accordingly, such that in each waveguide the amplitudes are moved away from box $0$ and towards box ${N-1}$ (the amplitude of the basis state with $\ket{1_{\mu,0}}$ moves to the state with $\ket{1_{\mu,1}}$, the amplitude of the state with $\ket{1_{\mu,1}}$ moves to the state with $\ket{1_{\mu,2}}$, etc.). Once the boxes are shifted in this way, the amplitudes of the states with $\ket{1_{\mu,0}}$ are set to zero, corresponding to a new empty box entering at the original position of box $0$ in both waveguides. This process does not conserve the norm of the state ${\ket{\psi(t+\Delta t)}}$, so the state must be renormalized at the end of the iteration in step~(5) before the next time step begins [as in Eq.~(\ref{eq:normalisation})].

Steps~(1)-(5) can be repeated to simulate the time evolution of the system until a desired end time is reached. Such a simulation is called a quantum trajectory. Since the simulation of losses in step~(2) and the measurement simulation in step~(3) are stochastic processes, an average of many trajectories must be obtained to compute the expected time evolution of observables, conditioned on photon detection events at the waveguide ends.


\newpage
\subsection{Trajectory Results}\label{subsec:trajectories_results}

In this section, we present example trajectory results for Device~1 and Device~2 from the main text. In all the results presented here, we use a time step ${\Delta t = 2 \times 10^{-7}}$~ns, and each waveguide is divided into ${N=2}$ boxes. Hence, the cavity-waveguide interaction takes place at box $0$, and photon number measurements are simulated at box $1$ in each waveguide (see Fig.~\ref{fig:discretized_waveguides}). We use a small number of waveguide boxes to keep the size of our matrices and vectors small in order to make the computations faster (the choice of $N$ does not affect our average trajectory results, as this effectively just determines the length of the waveguides). We also neglect losses in our trajectory simulations, i.e., we set ${\kappa_{\alpha} = \gamma_j = 0}$, and in step~(2) of the trajectory algorithm we always apply the time evolution operator $U_{\text{eff}}(\Delta t)$ at each time step [Eq.~(\ref{eq:no_loss})]. Neglecting losses does not have a significant effect on the directional contrast and allows us to speed up the computations.


\subsubsection{Device 1}

\begin{figure}[h!]
    \centering
    \includegraphics[width=0.5\textwidth]{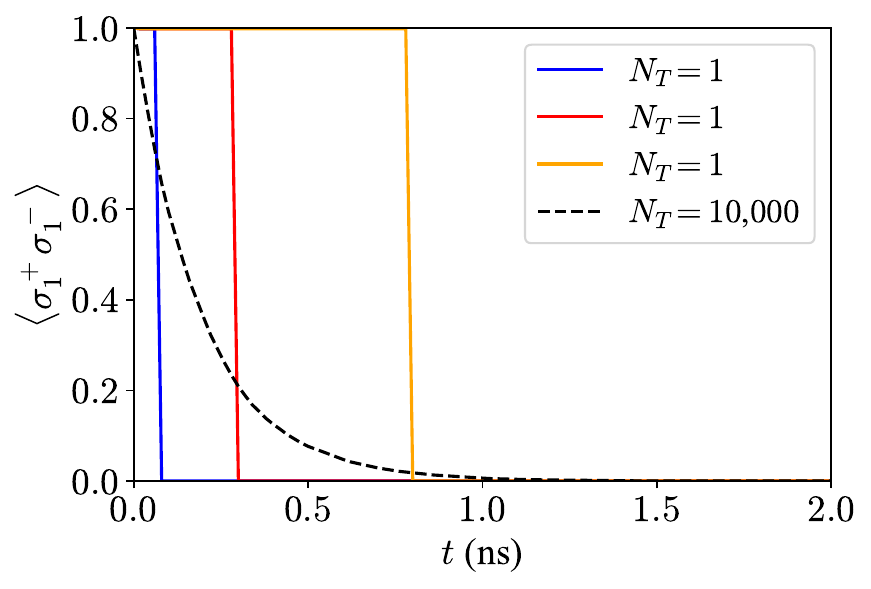}
    \caption{Example trajectories showing the excited state population ${\langle \sigma_1^+ \sigma_1^- \rangle}$ as a function of time $t$. The solid blue, red, and orange curves correspond to individual trajectory simulations, while the black, dashed curve is an average over $10\text{,}000$ trajectories.}
    \label{fig:example_trajectories}
\end{figure}

In Fig.~\ref{fig:example_trajectories}, we show example trajectories for the population ${\langle \sigma_1^+ \sigma_1^- \rangle}$ of the excited state $\ket{e_1}$ of the emitter in the cavity, for the parameters of Device~1 from the main text. Here, the cavity mode wavelengths are ${\lambda_{c,H} = 946.6}$~nm and ${\lambda_{c,V} = 948.9}$~nm, and the corresponding $Q$ factors are ${Q_H = 810}$ and ${Q_V = 320}$ (from which we obtain the cavity-waveguide coupling rates ${V_{R,\alpha} = V_{L,\alpha} = \omega_{c,\alpha}/2Q_{\alpha}}$ for ${\alpha \in \{H,V\}}$). Our measured data for Device~1 corresponds to a charged exciton, which we model as a circularly polarized two-level emitter [this represents one of the Zeeman components of the charged quantum dot (QD) in a magnetic field]. We do this by decoupling transition $2$ of the three-level system from the cavity by setting ${g_{H,2} = g_{V,2} = 0}$. The coupling between transition $1$ and the cavity is given by ${g_{H,1}/2\pi = 10}$~GHz and ${g_{V,1}/2\pi = 10i}$~GHz, which corresponds to right-handed ($\sigma^+$) circular polarization. For Fig.~\ref{fig:example_trajectories}, we use ${\lambda_{e,1} = 947}$~nm for the transition wavelength of the emitter ($\lambda_{e,2}$ is irrelevant as we decoupled transition $2$ from the cavity in this case).

To obtain the results in Fig.~\ref{fig:example_trajectories}, we start with an initial state where all the population is in the excited state $\ket{e_1}$ [Eq.~(\ref{eq:state}) with ${\alpha = 1}$ and ${\beta = 0}$], and calculate the expectation value ${\langle \sigma_1^+ \sigma_1^- \rangle}$ [Eq.~(\ref{eq:emitter_pop}) with ${j=1}$] at each time step. The solid blue, red, and orange curves correspond to individual trajectories (${N_T = 1}$, where we use $N_T$ to denote the number of trajectories). For these, we run steps~(1)-(5) of the algorithm in Section~\ref{subsec:trajectories_algorithm} until an end time of ${t_{\text{end}} = 2}$~ns. In these single trajectories, we observe jumps where the population of the excited state $\ket{e_1}$ falls to zero at a time when a photon is detected in one of the waveguides during the measurement simulations. This is because a photon detection event in one of the waveguides implies that the emitter must have decayed, so it is projected onto its ground state $\ket{g}$. The black, dashed curve corresponds to an average over $10\text{,}000$ of such stochastic trajectories, which accurately reproduces the exponential decay of the excited state population.

\begin{figure}[h!]
    \centering
    \includegraphics[width=0.7\textwidth]{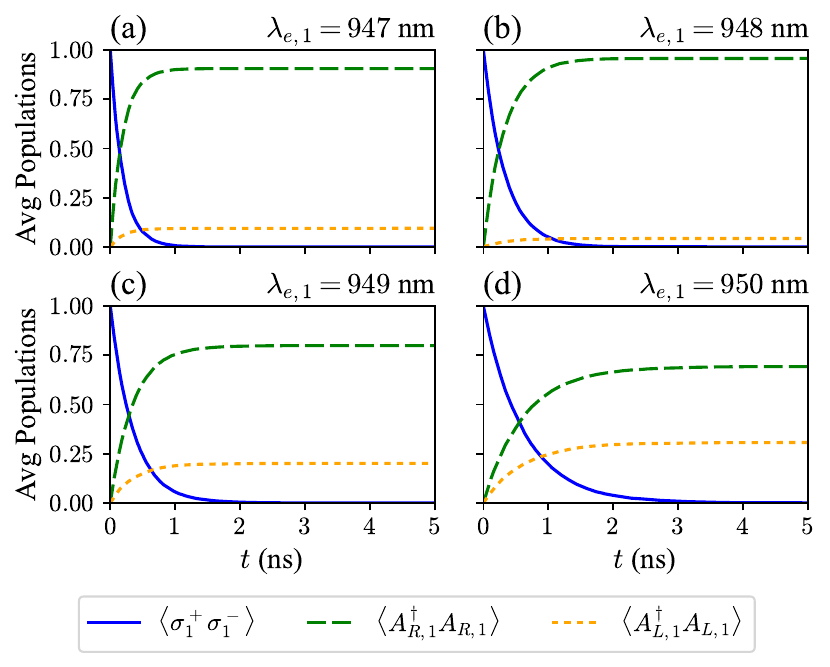}
    \caption{Average trajectory results for Device~1, where the transition wavelength of the two-level emitter is (a) ${\lambda_{e,1} = 947}$~nm, (b) ${\lambda_{e,1} = 948}$~nm, (c) ${\lambda_{e,1} = 949}$~nm, and (d) ${\lambda_{e,1} = 950}$~nm. The solid, blue curves correspond to the excited state population $\smash{\langle \sigma_1^+ \sigma_1^- \rangle}$, the dashed, green curves correspond to the population $\smash{\langle A_{R,1}^{\dag} A_{R,1}^{\vphantom{\dag}} \rangle}$ of the final box in the right waveguide, and the dashed, orange curves correspond to the population $\smash{\langle A_{L,1}^{\dag} A_{L,1}^{\vphantom{\dag}} \rangle}$ of the final box in the left waveguide. Each curve is an average over $10\text{,}000$ trajectory simulations.}
    \label{fig:device_1_trajectories}
\end{figure}

In Fig.~\ref{fig:device_1_trajectories}, we show average trajectories for different emitter wavelengths, where (a) ${\lambda_{e,1} = 947}$~nm (as in Fig.~\ref{fig:example_trajectories}), (b) ${\lambda_{e,1} = 948}$~nm, (c) ${\lambda_{e,1} = 949}$~nm, and (d) ${\lambda_{e,1} = 950}$~nm. We again consider a circularly polarized emitter, and use the same device parameters as in Fig.~\ref{fig:example_trajectories}. Here, the solid, blue curves correspond to the excited state population $\smash{\langle \sigma_1^+ \sigma_1^- \rangle}$ [Eq.~(\ref{eq:emitter_pop}), ${j=1}$], the dashed, green curves correspond to the population $\smash{\langle A_{R,1}^{\dag} A_{R,1}^{\vphantom{\dag}} \rangle}$ of the final box in the right waveguide [Eq.~(\ref{eq:wg_pop}), ${\mu = R}$ and ${N=2}$], and the dashed, orange curves correspond to the population $\smash{\langle A_{L,1}^{\dag} A_{L,1}^{\vphantom{\dag}} \rangle}$ of the final box in the left waveguide [Eq.~(\ref{eq:wg_pop}), ${\mu = L}$ and ${N=2}$]. We do not show the populations ${\langle c_{\alpha}^{\dag} c_{\alpha}^{\vphantom{\dag}} \rangle}$ of the cavity modes because they are negligible for all times $t$ (the cavity-waveguide coupling is much stronger than the emitter-cavity coupling, so the population is quickly transferred to the waveguides when the emitter decays). Each of the curves is an average over $10\text{,}000$ trajectory simulations. From the waveguide populations, we can extract the directional contrast $C$ using Eq.~(3) from the main text, with ${N=2}$ since we divide each waveguide into two boxes in these simulations:
\begin{equation}
C = \left| \frac{ \bigl\langle A_{R,1}^{\dag} A_{R,1}^{\vphantom{\dag}} \bigr\rangle_{t=t_{\text{end}}} - \bigl\langle A_{L,1}^{\dag} A_{L,1}^{\vphantom{\dag}} \bigr\rangle_{t=t_{\text{end}}} }{ \bigl\langle A_{R,1}^{\dag} A_{R,1}^{\vphantom{\dag}} \bigr\rangle_{t=t_{\text{end}}} + \bigl\langle A_{L,1}^{\dag} A_{L,1}^{\vphantom{\dag}} \bigr\rangle_{t=t_{\text{end}}} } \right|,
\label{eq:contrast}
\end{equation}
where ${t_{\text{end}} = 5}$~ns is the end time of the trajectories, chosen to ensure that the emitter had fully decayed to its ground state and all the photon population was transferred to the ends of the waveguides. The wavelength dependence of the directional contrast is clearly visible in Fig.~\ref{fig:device_1_trajectories}, where we can see that tuning the transition wavelength of the emitter changes the probability that the emitted photon ends up in the left and right waveguides. From these trajectory results, we obtain (a) ${C = 0.81}$, (b) ${C = 0.91}$, (c) ${C = 0.60}$, and (d) ${C = 0.39}$.

\begin{figure}[h!]
    \centering
    \includegraphics[width=0.5\textwidth]{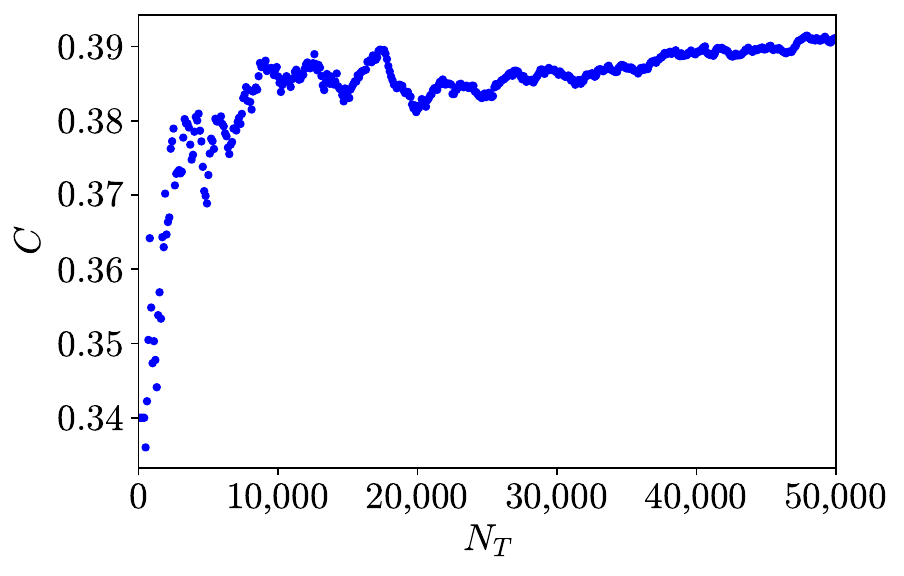}
    \caption{Contrast convergence test, showing the directional contrast $C$ as a function of the number of trajectories $N_T$ we average over when calculating the contrast from the waveguide populations using Eq.~(\ref{eq:contrast}).}
    \label{fig:contrast_convergence_test}
\end{figure}

In order to determine how many trajectories to average over to get an accurate result for the directional contrast $C$, we carried out a convergence test where we calculated the contrast as a function of the number of trajectories $N_T$. This is shown in Fig.~\ref{fig:contrast_convergence_test}, where we use the same parameters as for Fig.~\ref{fig:device_1_trajectories}(d). We calculate $C$ using Eq.~(\ref{eq:contrast}), where the waveguide populations are calculated by averaging over $N_T$ trajectories (in steps of $100$ up to ${N_T = 50\text{,}000}$). As expected, increasing $N_T$ reduces the uncertainty in the data points (for ${N_T > 10\text{,}000}$, this uncertainty is ${\ll 1\%}$). In Fig.~\ref{fig:contrast_convergence_test} we also see that the data points converge to a fluctuating curve, which arises from the additional layer of noise produced by the way in which the photon number measurement simulations are implemented with random numbers in Python. We choose ${N_T = 10\text{,}000}$ to obtain our contrast results, where there is a good balance between accuracy and simulation time.


\subsubsection{Device 2}

\begin{figure}[h!]
    \centering
    \includegraphics[width=0.75\textwidth]{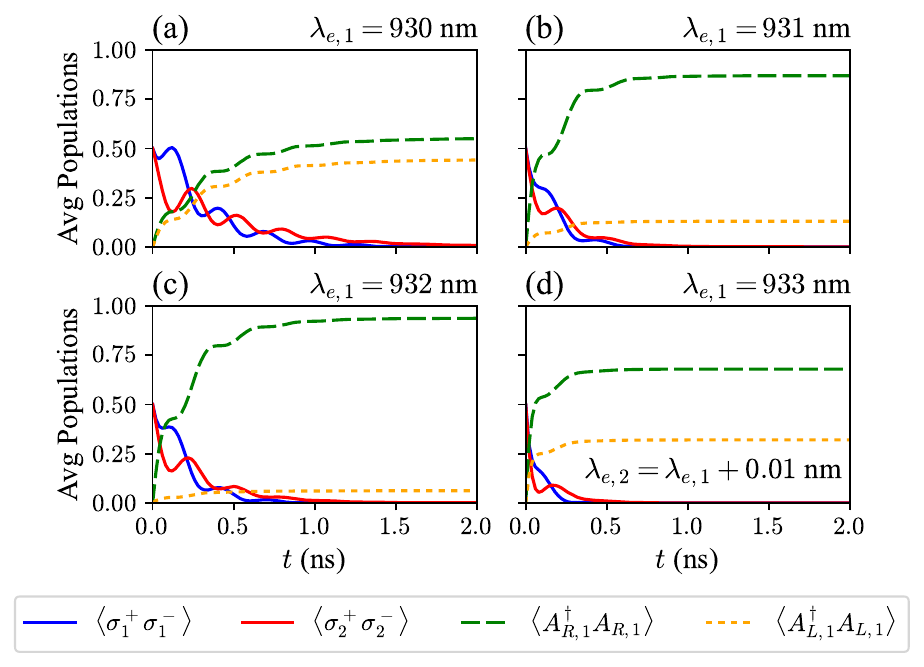}
    \caption{Average trajectory results for Device~2, where we consider a three-level emitter with (a) ${\lambda_{e,1} = 930}$~nm, (b) ${\lambda_{e,1} = 931}$~nm, (c) ${\lambda_{e,1} = 932}$~nm, and (d) ${\lambda_{e,1} = 933}$~nm, and the second transition has wavelength ${\lambda_{e,2} = \lambda_{e,1} + 0.01}$~nm (corresponding to a fine structure splitting of $0.01$~nm). The solid blue and red curves correspond to the excited state populations ${\langle \sigma_1^+ \sigma_1^- \rangle}$ and ${\langle \sigma_2^+ \sigma_2^- \rangle}$, respectively. The dashed, green curves correspond to the population $\smash{\langle A_{R,1}^{\dag} A_{R,1}^{\vphantom{\dag}} \rangle}$ of the final box in the right waveguide, and the dashed, orange curves correspond to the population $\smash{\langle A_{L,1}^{\dag} A_{L,1}^{\vphantom{\dag}} \rangle}$ of the final box in the left waveguide.}
    \label{fig:device_2_trajectories}
\end{figure}

In Fig.~\ref{fig:device_2_trajectories}, we show averaged trajectories for the parameters corresponding to Device~2 from the main text. Here, the cavity mode wavelengths are ${\lambda_{c,H} = 930.9}$~nm and ${\lambda_{c,V} = 933.6}$~nm, and the corresponding $Q$ factors are ${Q_H = 660}$ and ${Q_V = 779}$. The measured directional contrast in Device~2 corresponds to neutral exciton QD states, which we model as a three-level emitter with a fine structure splitting of $0.01$~nm (this corresponds to one of the Zeeman components of a neutral exciton in a magnetic field). A splitting of $0.01$~nm is a typical order-of-magnitude value for self-assembled InAs QDs~\cite{Mar2016}. The asymmetry of the measured QD that causes the fine structure splitting also modifies the emitter-cavity coupling, and hence the polarization of the emitted photon. For Fig.~\ref{fig:device_2_trajectories}, we use ${g_{H,1}/2\pi = g_{H,2}/2\pi = 10}$~GHz and ${g_{V,1}/2\pi = g_{V,2}/2\pi = 20 e^{5\pi i/8}}$~GHz for the emitter-cavity coupling rates. We use a small ($\pi/8$) phase rotation away from the $\pi/2$ phase that would be present for a perfect circularly polarized emitter, which is expected to arise due to the asymmetry of the QD~\cite{Bayer2002} (e.g., if the transition dipole moment is no longer oriented along the direction of the applied magnetic field). We also use unequal coupling to the $H$ and $V$ modes of the cavity (${|g_{V,j}| = 2 |g_{H,j}|}$), which can arise from the location of the QD in the cavity and the spatial profile of the cavity mode fields~\cite{ChiralCav_2022} [see Fig.~3(d) in the main paper, for example].

The trajectory results shown in Fig.~\ref{fig:device_2_trajectories} correspond to the case where the wavelength of transition $1$ of the emitter is (a) ${\lambda_{e,1} = 930}$~nm, (b) ${\lambda_{e,1} = 931}$~nm, (c) ${\lambda_{e,1} = 932}$~nm, and (d) ${\lambda_{e,1} = 933}$~nm, and ${\lambda_{e,2} = \lambda_{e,1} + 0.01}$~nm. As we use a non-resonant excitation scheme in our experimental measurements, we expect that the two excited states $\ket{e_1}$ and $\ket{e_2}$ of each neutral exciton Zeeman component are populated equally, so our initial state for the trajectory simulations corresponding to Device~2 is an equal superposition of $\ket{e_1}$ and $\ket{e_2}$ [Eq.~(\ref{eq:state}) with ${\alpha = \beta = 1/\sqrt{2}}$]. The solid blue and red curves show the time evolution of the populations ${\langle \sigma_1^+ \sigma_1^- \rangle}$ and ${\langle \sigma_2^+ \sigma_2^- \rangle}$ of excited states $1$ and $2$, respectively. As in Fig.~\ref{fig:device_1_trajectories}, the dashed, green curves correspond to the population ${\langle A_{R,1}^{\dag} A_{R,1}^{\vphantom{\dag}} \rangle}$ of the final box in the right waveguide, and the dashed, orange curves correspond to the population $\smash{\langle A_{L,1}^{\dag} A_{L,1}^{\vphantom{\dag}} \rangle}$ of the final box in the left waveguide. Each curve is an average over $10\text{,}000$ trajectory simulations. We use an end time of ${t_{\text{end}} = 5}$~ns for all the trajectories to ensure that the excited state populations had fully decayed, but we only show the first two nanoseconds in Fig.~\ref{fig:device_2_trajectories} to more clearly visualize the excited state population dynamics at short times. We observe oscillations in the populations due to interference corresponding to coherent population transfer between the two excited states of the emitter through the common ground state. From the averaged waveguide populations, we calculate the directional contrast using Eq.~(\ref{eq:contrast}), and find that (a) ${C = 0.11}$, (b) ${C = 0.74}$, (c) ${C = 0.88}$, and (d) ${C = 0.36}$.


\newpage
\section{Expanded Results for Asymmetric Directional Contrast}

In this section, we present detailed trajectory results for varying polarization states of QD transitions, corresponding to the asymmetric directional contrast in the cavity-waveguide system. We analyze how deviations in polarization states associated with different emitter-cavity coupling rates $g_{\alpha,j}$ influence the directional contrast $C$ and its wavelength dependence. 

\begin{figure}[h]
    \centering
    \includegraphics[width=0.95\textwidth]{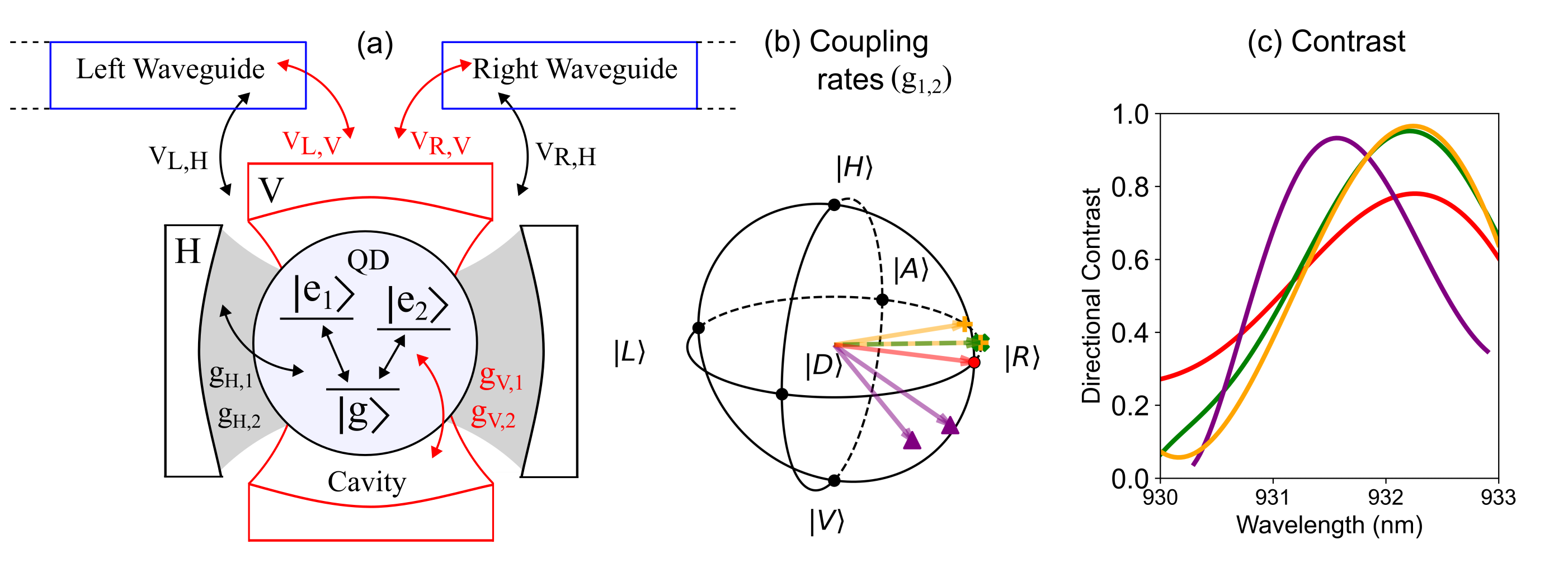}
    \caption{(a) Schematic of the waveguide-coupled nanocavity. Transition $j \in \{1,2\}$ of the QD ($\ket{g} \leftrightarrow \ket{e_j}$) has frequency $\omega_{e,j}$ and couples to cavity mode $\alpha \in \{H,V\}$ (having resonance frequency $\omega_{c,\alpha}$) with coupling rate $g_{\alpha,j}$. Cavity mode $\alpha$ also couples to the right and left waveguides with coupling rates $V_{R,\alpha}$ and $V_{L,\alpha}$, respectively. (b) Bloch sphere representation of different polarization states, corresponding to varying emitter-cavity coupling rates $g_{\alpha,j}$ in the directional cavity system. (c) Directional contrast for the emitter-cavity coupling rates used in (b), calculated using the trajectories model [Eq.~(\ref{eq:contrast})].}
    \label{fig:contrast_bloch}
\end{figure}

Examples of different polarization states of QD transitions are represented on the Bloch sphere in Fig.~\ref{fig:contrast_bloch}(a). These states correspond to specific coupling rates $g_{\alpha,j}$, which determine the emission characteristics of the system. The Bloch sphere uses a coordinate system $(x, y, z)$ to describe the polarization state, where $z$ represents the population difference between two basis states (e.g., right- and left-circular polarization), and $x$ and $y$ correspond to the in-phase and out-of-phase coherence between the states, respectively. The results in Fig.~\ref{fig:contrast_bloch}(b) highlight how modifications to the polarization states affect the directional contrast. It is important to note that we are not suggesting that the QDs themselves are inherently elliptically polarized; rather, these effective couplings arise due to factors such as emitter misalignment within the cavity, fine structure splitting, and relative differences in the Purcell enhancement for H and V polarizations. In order to demonstrate how these deviations can cause changes in the predicted contrast, we explore the case of the $\sigma^{+}$ transition of Device~2 in more detail:

\begin{itemize}
    \item \textbf{Red: Circular Polarization Coupling (0.00, 1.00, 0.00)}: Both transitions are circularly polarized, yielding a baseline directional contrast that is equivalent to the degree of circular polarization at the cavity center [see Fig.~5(d) in the main text]. The contrast remains broad and insensitive to small wavelength shifts.

    \item \textbf{Green: Elliptical Polarization Coupling (-0.38, 0.92, 0.00)}: Deviations from pure circular polarization increase the peak contrast. 

    \item \textbf{Yellow: Asymmetric Polarization Coupling, Transition 1: (-0.31, 0.74, -0.60); Transition 2: (-0.71, 0.71, 0.00)}: Here we allow the two fine structure transitions of the three-level system modeling a neutral exciton Zeeman component to have different coupling rates. The transitions are separated by a fine structure splitting of $0.01$~nm, or about $15$~$\mu$eV. Although the maximum achievable contrast remains unchanged, the contrast curve narrows in comparison to Green. This narrowing indicates a heightened sensitivity to small wavelength shifts at shorter wavelengths.
    
    \item \textbf{Purple: Asymmetric Polarization Coupling and Anisotropic Purcell Enhancement, Transition 1: (-0.36, 0.71, -0.60); Transition 2: (-0.46, 0.39, -0.80)}: Similarly to Yellow, for the two transitions of the modeled neutral exciton, different couplings are given to the cavity modes, with the addition of a greater magnitude to the coupling strength to the vertical ($V$) mode over the horizontal ($H$) mode, modeling points in the cavity where the electric field concentration may be greater for the vertical mode over the horizontal. Here, the contrast curve has a narrower wavelength dependence, and blue shifts. These are the parameters used for FIG \ref{fig:DeviceTwo}. 
\end{itemize}

The parameters used are the following.
\begin{itemize}
    \item \textbf{Emitter-Cavity Coupling Rates:} \(g_{H,1}/2\pi = 10\)~GHz and \(g_{V,1}/2\pi = 10i\)~GHz for circular (\(\sigma^+\)) polarization.
    \item \textbf{Cavity Parameters:} \(\lambda_{c,H} = 930.9\)~nm, \(\lambda_{c,V} = 933.6\)~nm, \(Q_H = 660\), \(Q_V = 779\).
    \item \textbf{Simulation Conditions:} Time step \(\Delta t = 2 \times 10^{-7}\)~ns, waveguides divided into \(N=2\) boxes each. Losses are neglected (\(\kappa_{\alpha} = \gamma_j = 0\)).
\end{itemize}


\newpage
\section{Waveguide Coupling and Crosstalk}

\begin{figure}[h!]
    \centering
    \includegraphics[width=0.99\textwidth]{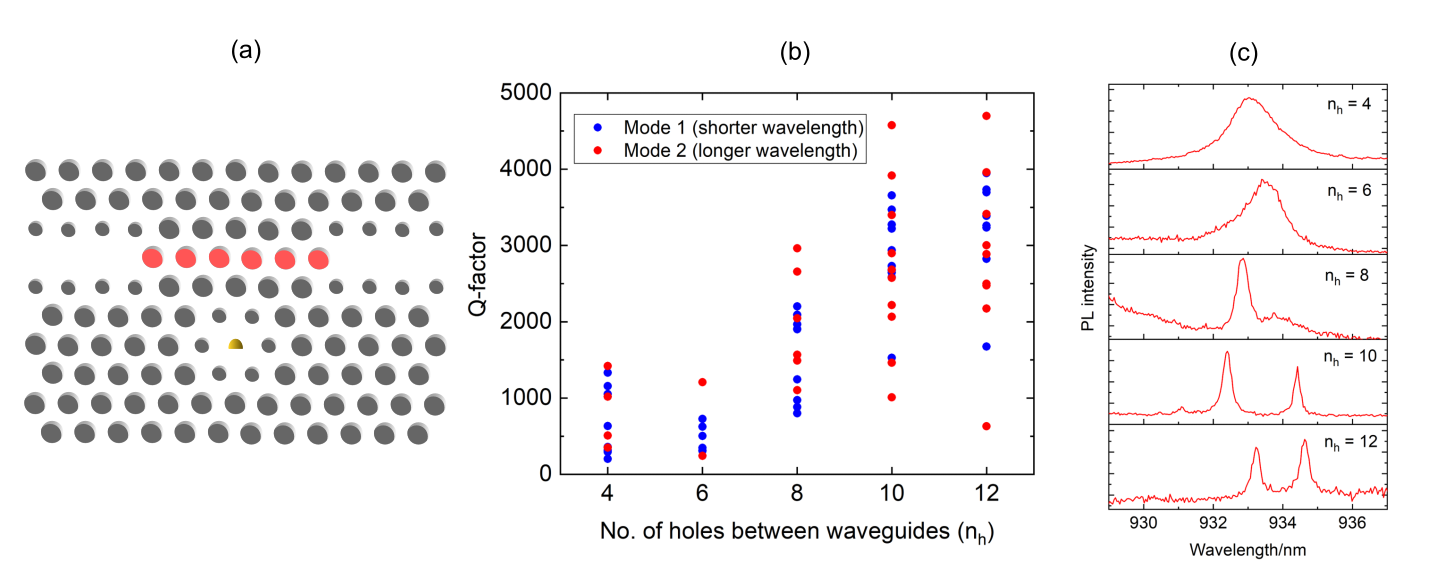}
    \caption{(a) Schematic of the device with the apertures between the waveguides highlighted in red. (b) The measured $Q$ factors of different cavity devices in relation to the number of holes $n_{h}$ between the waveguides. As the number of holes between the waveguides increases, so does the mean measured $Q$ factor. (c) PL spectra of the cavity modes for devices with equal air hole radii and different values of $n_{h}$. The $Q$ factor of the modes can clearly be seen to reduce as $n_{h}$ decreases. The wavelengths of the cavity modes remain roughly constant. At low values of $n_{h}$ the two cavity modes begin to overlap~\cite{Brunswick}.}
    \label{fig:holes}
\end{figure}

Fig.~\ref{fig:holes}(b) shows the effect of changing the number of holes $n_{h}$ between the two waveguides on the experimentally measured $Q$ factors of the cavity modes [these holes are highlighted in red in Fig.~\ref{fig:holes}(a)]. A decrease in $n_{h}$ results in a lower $Q$ factor due to enhanced cavity-waveguide coupling, a consequence of the reduced separation. This is more pronounced at smaller values of $n_{h}$. When $n_{h}$ is $10$ or more, the mean $Q$ factor reaches a stable value, similar to an uncoupled cavity. Interestingly, the lowest average $Q$ factor is observed when ${n_{h} = 6}$, not at the smallest $n_{h}$ of $4$, likely due to optimal mode coupling at this value. Fig.~\ref{fig:holes}(c) presents the photoluminescence (PL) spectra corresponding to data from (b), comparing devices with the same hole radii but different $n_{h}$. Again, the $Q$ factor can be seen to decrease when $n_h$ is reduced. The two cavity modes also start to overlap at small values of $n_h$~\cite{Brunswick}. It is possible to fine-tune the mode waist radius of a photonic crystal, which could provide a way to further finely control the coupling between the waveguides and the cavity~\cite{taper}.


\newpage
\section{Wafer Details}

The wafer used to fabricate the devices is grown via molecular beam epitaxy. A GaAs substrate is used, on which a $1.05\,\mu$m AlGaAs sacrificial layer is grown. The $170$~nm GaAs membrane is then grown on top of this AlGaAs layer. The membrane is engineered to form a p-i-n diode, with $30$~nm p-doped GaAs at the top and $30$~nm n-doped GaAs at the bottom, sandwiching an intrinsic region containing a layer of InAs QDs. AlGaAs tunneling barriers are incorporated on either side of the QD layer to improve charge confinement.

After growth, the AlGaAs sacrificial layer is selectively etched to produce a free-standing GaAs membrane. The photonic crystal pattern is then defined by electron beam lithography and transferred into the membrane via dry etching, creating the triangular lattice of air holes and the H1 cavity and W1 waveguides. The resulting structure supports high-quality optical modes in the desired wavelength range, enabling efficient coupling to and from the embedded QDs.


\newpage
\section{Relationship between Left- and Right-Circular Polarization Coefficients under Misalignment}
\label{sec:supplementary_misalignment}

In this section, we discuss the effect of emitter misalignment on the complex polarization coefficients corresponding to left‐ and right‐circularly polarized transitions of a QD.


\subsection{Circular Polarization States}

An ideal QD emits photons in pure right‐circular ($\ket{R}$) or left‐circular ($\ket{L}$) polarization states. In the absence of misalignment, these states can be written in terms of the horizontal ($\ket{H}$) and vertical ($\ket{V}$) linear polarization basis as:
\begin{align}
\ket{R} &= \frac{1}{\sqrt{2}} \Big( \ket{H} + i\,\ket{V} \Big), \\
\ket{L} &= \frac{1}{\sqrt{2}} \Big( \ket{H} - i\,\ket{V} \Big).
\end{align}
The coefficients for $\ket{L}$ are the complex conjugates of those for $\ket{R}$. Explicitly, if we write
\[
\begin{aligned}
\ket{R} &= \; c_H^R\,\ket{H} + c_V^R\,\ket{V}, 
\end{aligned}
\]
then
\[
\begin{aligned}
\ket{L} &= \; c_H^L\,\ket{H} + c_V^L\,\ket{V}, 
\end{aligned}
\]
with the relationships
\begin{align}
c_H^L &= (c_H^R)^*, \\
c_V^L &= (c_V^R)^*.
\end{align}


\subsection{Effect of Real 3D Rotations (Misalignment)}

When the emitter or cavity is misaligned relative to the lab frame, the state vectors undergo a real three-dimensional rotation described by a rotation matrix \(R\). Importantly, such a rotation is a linear transformation that does not introduce additional complex phases. Instead, it mixes the components of the polarization vector while preserving conjugate relationships.

Under this real rotation, the coefficients for the right‐circular state transform as
\[
\begin{pmatrix} 
c_H^R \\ 
c_V^R 
\end{pmatrix}
\mapsto 
\begin{pmatrix} 
c_H^{R'} \\ 
c_V^{R'} 
\end{pmatrix}
= 
R \begin{pmatrix} 
c_H^R \\ 
c_V^R 
\end{pmatrix},
\]
where \(c_H^{R'}\) and \(c_V^{R'}\) are the rotated coefficients. Since \(R\) is a real matrix, the relationship between the left‐ and right‐circular polarization coefficients remains preserved under the transformation:
\begin{align}
c_H^{L'} &= (c_H^{R'})^*, \\
c_V^{L'} &= (c_V^{R'})^*,
\end{align}
where \(c_\alpha^{L'}\) and \(c_\alpha^{R'}\) (\(\alpha \in \{H,V\}\)) are the coefficients for the rotated \(\ket{L}\) and \(\ket{R}\) states, respectively.


\subsection{Impact of Quantum Dot Symmetry on Polarization}

The above analysis assumes that the QD emits pure circularly polarized light under ideal conditions. However, imperfections in the symmetry of the QD can alter its fine structure and the polarization properties of its transitions. Such symmetry deviations may lead to:
\begin{itemize}
    \item Mixed polarization states, where the emitted light is elliptical rather than purely circular.
    \item Variations in the intrinsic dipole orientations of the emitter transitions, affecting how they couple to the cavity modes.
\end{itemize}
When these symmetry imperfections are significant, the straightforward complex conjugate relationship between the \(\ket{L}\) and \(\ket{R}\) coefficients may no longer hold perfectly. Instead, deviations from pure circular polarization introduce additional complexities in the coupling rates \(g_{\alpha,j}\) and the observed directional emission characteristics. Accurate modeling in such cases requires accounting for the modified polarization states and potentially nonideal phase relationships arising from the QD's broken symmetry. Nonetheless, the core impact of real rotations on preserving conjugate relationships provides a foundational understanding, which can be adjusted to incorporate symmetry-induced polarization effects when necessary.

\end{document}